\documentclass[5p,authoryear]{elsarticle}

\usepackage{hyperref}

\usepackage{amsfonts}
\usepackage{amsmath}
\usepackage{amssymb}
\usepackage{booktabs}

\usepackage[figuresright]{rotating}

\journal{Astronomy \& Computing}

\begin{document}

\begin{frontmatter}

\title{Using Multiple Instance Learning for Explainable Solar Flare Prediction}

\author{C. Huwyler\fnref{email}}
\fntext[email]{Email: cedric.huwyler@fhnw.ch}

\author{M. Melchior}

\address{Institute for Data Science, Fachhochschule Nordwestschweiz (FHNW)\\Bahnhofstrasse 6, 5210 Windisch\\Switzerland}

\begin{abstract}
In this work we leverage a weakly-labeled dataset of spectral data from NASA’s IRIS satellite for the prediction of solar flares using the Multiple Instance Learning (MIL) paradigm. While standard supervised learning models expect a label for every instance, MIL relaxes this and only considers bags of instances to be labeled. This is ideally suited for flare prediction with IRIS data that consists of time series of bags of UV spectra measured along the instrument slit. In particular, we consider the readout window around the Mg II h\&k lines that encodes information on the dynamics of the solar chromosphere. Our MIL models are not only able to predict whether flares occur within the next $\sim$25 minutes with accuracies of around $90\%$, but are also able to explain which spectral profiles were particularly important for their bag-level prediction. This information can be used to highlight regions of interest in ongoing IRIS observations in real-time and to identify candidates for typical flare precursor spectral profiles. We use k-means clustering to extract groups of spectral profiles that appear relevant for flare prediction. The recovered groups show high intensity, triplet red wing emission and single-peaked h and k lines, as found by previous works. They seem to be related to small-scale explosive events that have been reported to occur tens of minutes before a flare. 
\end{abstract}


\end{frontmatter}

\section{Introduction\label{sec:intro}}

The Multiple Instance Learning (MIL) paradigm was originated from drug activity prediction problems \citep{dietterich1997} and has recently gathered attention in the domain of image segmentation (e.g. \citet{oquab2015, pinheiro2015}) and especially in medical imaging (e.g. \cite{wang2017, sadafi2020, liu2018, sudharshan2019, kanavati2021}). In the MIL setting, label information is not available for single instances, but only for groups or so called \textit{bags} of multiple instances. As such, MIL is part of the weakly supervised learning domain \citep{zhou2018}. The relationship between individual instances and the overall bag label is made explicit by modeling how instance-level information is aggregated to bag-level information \citep{carbonneau2019, amores2013, fouldsfrank2010}. This allows MIL models to use the available bag-level label information to reveal the hidden instance-level labels. In the image segmentation domain, a pixel-based segmentation map is learned only from an image-level label.
Recently, it has been proposed that an attention-based MIL model shows supreme bag-level performance and allows to easily produce saliency maps that show on which instances the model focuses to make a particular bag-level prediction \citep{abmil}.

Physics-based modeling of observed phenomena is usually done in a bottom-up fashion, where the interactions between individual microscopic instances are modeled and then aggregated to a macroscopic quantity. The results on the macroscopic level are then compared to actual observations. In contrast, data-driven approaches are usually only able to model physical phenomena in a top-down fashion, where the macroscopic quantities (such as bag labels) are known but no information about the states of the microscopic quantities (individual instances) is available. Often, data-driven approaches show difficulties in revealing how they consult the available instances to reach their bag-level prediction. This frequently hinders such \textit{black box} models to get accepted in the broader research community. To this end MIL is a very attractive candidate, since it makes the relationship between bags and instances manifest, and is able to discover key instances in the decision process of the model.

In this work we apply the MIL paradigm to the problem of solar flare prediction. Solar flares are sudden and intense bursts of radiation from the Sun across the entire electromagnetic spectrum that are created by rapid downflow and subsequent deceleration of hot plasma triggered by reconnection of magnetic loops \citep{benz2016}. They are often (but not necessarily) accompanied by the ejection of plasma into space through the solar corona (\textit{coronal mass ejection}) and  occur mainly in so-called \textit{active regions} of the Sun. The magnitude of solar flares is conventionally indicated by the peak flux of the X-ray radiation emitted at wavelengths between 1-8 \AA{} and grouped into the classes X, M, C, B and A with the peak flux decreasing on a logarithmic scale. The high amounts of radiation and charged particles from strong solar flares and coronal mass ejections interact with Earth's magnetosphere. This interaction can lead to severe geomagnetic storms that last for several hours and can have widespread effects on terrestrial infrastructure and the near-Earth space environment \citep{lanzerotti2007, pulkkinen2007}. The ability to predict strong solar flares of class X and M tens of minutes if not hours in advance would yield valuable preparation time.

The prediction of solar flares is an active topic in the space weather community \citep{ahmed2013, bobra2015, boucheron2015, nishizuka2017, liu2017, jonas2018, florios2018, huang2018, liu2019, armstrong2019, galvez2019, chen2019, ahmadzadeh2021}. To this end, solar data recorded in different wavelengths and spatio-temporal resolutions are gathered by a fleet of satellites and an array of telescopes. Prediction of solar flares was so far mainly undertaken with data from the \textit{solar dynamics observatory} (SDO) \citep{pesnell2012} that continually observes the full solar disk at different ultraviolet (UV) passbands. Various levels of success are reported that are often difficult to compare because they are based on different pre-processing pipelines, different training and validation set splits and different performance metrics.

\begin{figure}[ht!]
  \begin{center}
  \includegraphics[width=\linewidth]{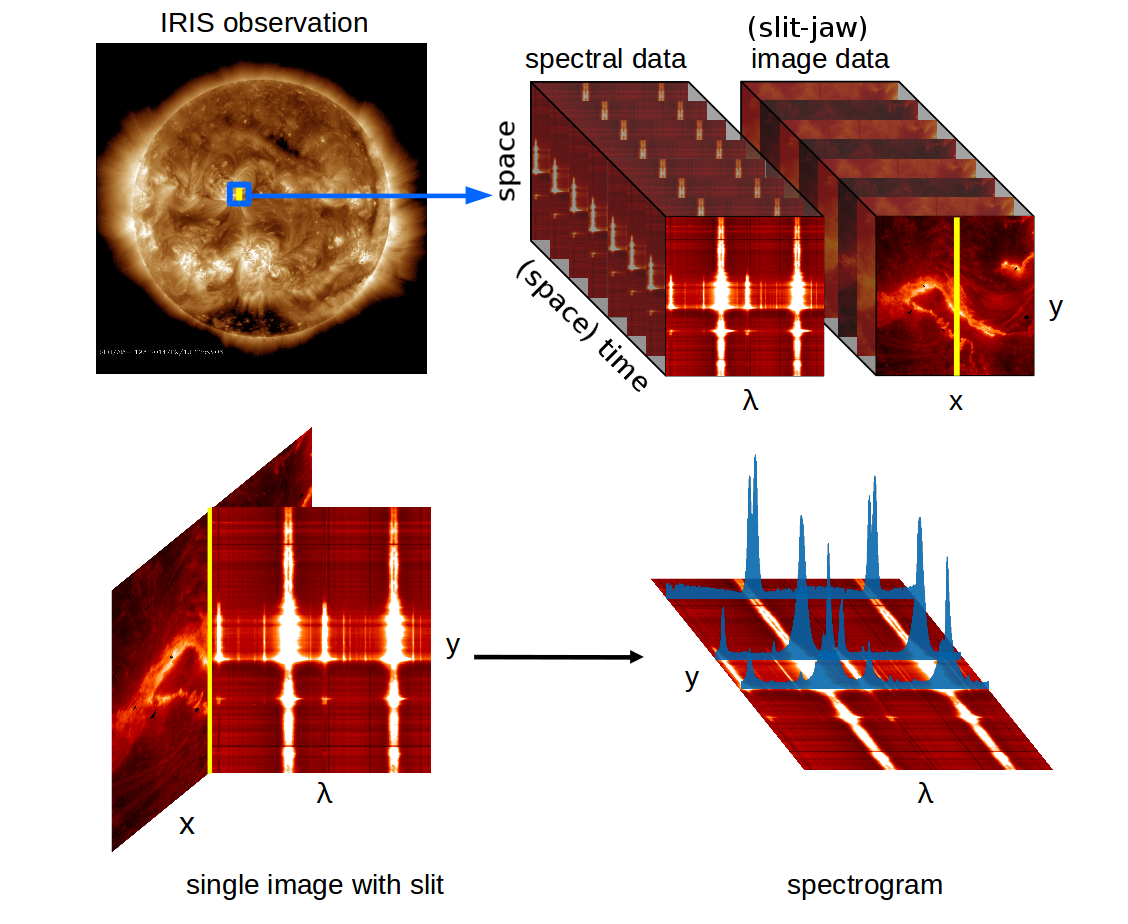}  
  \end{center}
  \caption{How IRIS records data.\newline\textbf{Top:} IRIS observes time series of small particular high-resolution patches of the solar surface according to a predefined observation schedule (left). It provides us with a spatio-temporal datacube of spectrograms along the slit accompanied by a data cube of images of the region around the slit (right).
  \newline 
  \textbf{Bottom:} Spectrograms are only measured along the position of the slit (yellow) in the center of the recorded image (left). A spectrogram contains a spectrum for each $y$-pixel along the slit (right). 
  }
  \label{fig:iris_overview}
\end{figure}

\begin{figure}[ht!]
  \begin{center}
  \includegraphics[width=\linewidth]{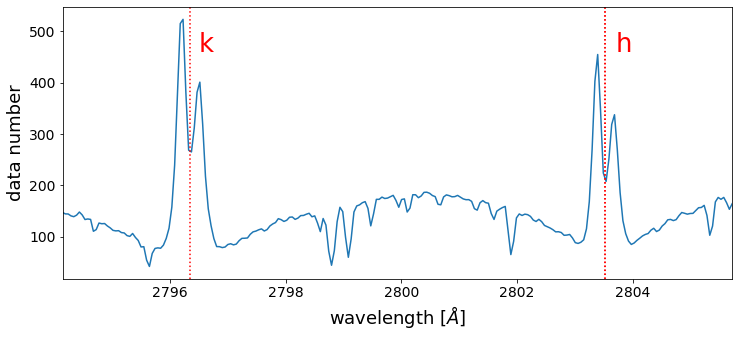}  
  \end{center}
  \caption{Typical example of a quiet-sun spectral profile in the Mg II readout window. The Mg II k and h lines are clearly visible and show the typical central core reversal (indicated with the red-dashed lines).}
  \label{fig:mgII_example}
\end{figure}

We focus in this work on high-resolution spectral data from the near-ultraviolet regime, specifically on a readout window around the Magnesium (Mg) II k and h line cores centered at 2796.34 and 2803.52 \AA{} (Fig.~\ref{fig:mgII_example} shows an example of a quiet-sun spectral profile). These data were collected by NASA's \textit{Interface Region Imaging Spectrograph} (IRIS) satellite \citep{irismissionpaper} that is designed to observe the transition region between the solar chromosphere and the corona, and it provides an unprecedented spectral (53 m\AA), spatial (0.33-0.4 arcsec) and temporal resolution (two-second cadence) along a vertical slit pointed onto a small patch on the Sun. In addition, also high-resolution images of the region around the slit (\textit{slit-jaw images, SJI}) in up to four wavebands are recorded (Fig.~\ref{fig:iris_overview}).
The Mg II h\&k window encodes information on the dynamics of the entire chromosphere \citep{vernazza1981, leenaarts2013_1, leenaarts2013_2, leenaarts2013_3}, such as plasma velocities, temperatures and formation heights. A recent study found a set of unique spectral shapes in this window that clearly indicate the presence of a flare \citep{panos2018}. A further study based on similar data has used specific expert-based features to show that flare prediction in the Mg II h\&k readout window could in principle be possible within a 25 minute window \citep{panoskleint2020}. In particular, a decreasing ratio between the Mg k and h integrated intensities seems to indicate an upcoming flare already before its impulsive phase. 

While spectral data from IRIS comes at an unprecedented spatio-temporal resolution in the UV regime, considerable pre-processing is required to bring the dataset in a form suitable for machine learning. In particular, three main challenges render the application of standard supervised machine learning models difficult:

\paragraph{Weak labels}

When working with IRIS data, the individual spectra seem to provide a natural fundamental unit of data. However, labels are not available on the level of individual spectra, but only on the level of entire observations (e.g. non-flaring active region, pre-flare or flare). Although some spectral shapes have been related to certain classes (e.g. according to  \cite{panos2018} some are clear indicators for a flare, however they do not appear in all flares), it is generally unclear how individual spectra are responsible for the overall label of an observation.

\paragraph{Different Observation Settings}

In contrast to SDO that continually observes the full solar disk, IRIS focuses on a small field-of-view (FOV) of up to 175 arcsec $\times$ 175 arcsec (see Fig.~\ref{fig:iris_overview}) that is defined according to an observation schedule. Limited by IRIS' downlink rate, a trade-off between spatial, temporal and spectral resolution has to be made by the observer, resulting in different settings for each observation. In addition, IRIS operates in different observing modes (fixed \textit{sit-and-stare} or moving \textit{n-step raster}), as introduced in section \ref{sec:preproc}.

\paragraph{Class imbalance}

The typical class imbalance to encounter in observational data from the Sun is even more pronounced. Only 10 X-class and 57 M-class flares happened to be observed by IRIS over the last solar cycle\footnote{According to the IRIS flare list distributed on \url{iris.lmsal.com}}. A significant part of these flares do not qualify for our dataset (see criteria in section \ref{sec:dataset}), reducing the available observations to 6 X-class and 13 M-class flares. This is a significantly low number in contrast to the total number of $\sim 30000$ observations since the commissioning of IRIS in 2013 and the total number of observed active regions that we estimate to be in the hundreds. 

\hspace{0.2cm}

In this work we address the problem of solar flare prediction given only weak labels using the MIL paradigm. To this end we treat each timestep of our 25 minute long observation time windows as an unsorted bag of individual spectra, thereby ignoring in this work any spatial structure (correlations between neighboring spectra) or the temporal dynamics (correlations between subsequent spectra) of the problem. This allows us to assign bag labels to active regions that are flaring or non-flaring in the near future. The resulting dataset can then be used to learn prediction models and to analyze what groups of spectral profiles dominate the decision process. In addition, the chosen approach uses a pooling function which allows to deal with differently sized bags of spectra. To our knowledge, we are the first to apply the MIL paradigm to the solar flare prediction problem.

\section{Dataset\label{sec:dataset}}

\subsection{Selection of sample observations}

Our dataset consists of 37 IRIS observations labeled either as (non-flaring) active region (AR) or pre-flare (PF) active region (listed in tables \ref{tab:arobs} and \ref{tab:pfobs}) with tens to hundreds of timesteps. 
Selecting representative observations for each class is a challenge because they need to satisfy a range of criteria: the FOV and especially the vertical slit should be centered on events that are characteristic for the class, the measured data should contain the Mg II h\&k readout window and they should not be located too close to the solar limb\footnote{Since spectra are the superposition of radiation absorbing and emitting layers of the solar atmosphere, spectra around the solar center are built up from radiation observed mostly in the vertical direction with respect to the solar surface, while spectra on the solar limb are built up from radiation observed more horizontally and show different characteristics.}. These criteria are only satisfied by 6 X-class flares and 13 M-class flares, resulting in a total of 19 PF observations (flare start times are indicated in table \ref{tab:flare_start_times}). To mitigate class imbalance, we selected a representative sample of 18 AR observations that do not flare within the next couple of hours, identical to the one used by \citet{panoskleint2020}. Also similar to \citet{panoskleint2020}, 25 minutes from each observation that are representative for the given class have manually been extracted, i.e. every timestep of the observation should encode some information about whether there is going to be a flare or not within the next 25 minutes. \citet{panoskleint2020} justify the small time window of 25 minutes by the fact that IRIS observed on average only 25 minutes before each major flare. The assumption that each timestep has the same label as the whole selected observation time window allows us to broadcast the label of each observation (AR or PF) to each of its timesteps and enables us to expand the 37 labeled observations to roughly 10'000 labeled bags containing roughly $600$ instances on average.

\subsection{Data pre-processing\label{sec:preproc}}

As introduced, IRIS' distinguishing feature is its spectrograph that refracts light along a vertical slit into its different wavelength components. The vertical slit (indicated in Fig.~\ref{fig:iris_overview} with a yellow line) along which spectral data is measured is typically composed of a few hundred vertically aligned pixels. IRIS operates in two different observation modes: fixed \textit{sit-and-stare} (always observing the same patch, co-moving with the rotation of the Sun) and moving \textit{n-step raster} (sweeping over a larger region of the Sun in a repeated manner). Hence, IRIS spectral data comes in the form of spatio-temporal data cubes that contain one time axis (for sit-and-stare) or space-time axis (for n-step raster) $t$, the $y$-axis as a spatial axis (along the vertical slit) and a spectral axis $\lambda$.

All selected observations in tables \ref{tab:arobs} and \ref{tab:pfobs} include the same spectral readout window around the Mg II h and k line cores. As the exact boundaries of the Mg II h\&k readout window differ from observation to observation, we extracted the specific window between 2794.14-2805.72 \AA{} that is contained in all observations and linearly interpolated the available spectral bins to a common length of 240 equally-spaced bins. We removed spectra that contain missing values (at least one value $<-100$) or are underexposed (less than 10 counts per bin) and clipped spectra with negative values to zero. Furthermore, we normalized the spectral profiles to the range $[0,1]$ and hence let our models focus more on the shape of spectra and on relative intensity differences. Note that the normalized profiles still provide access to a proxy of the absolute intensities of the Mg II h\&k peaks through the scaled intensity of the continuum radiation between the two peaks. The h and k peak intensities are strong predictors since they are strongly correlated with chromospheric gas temperatures \citep{kerr2015}.

The selected and pre-processed slices are stored in three-dimensional spatio-temporal data cubes with dimensions $[t,y,s]$, where $t$ is a time-step index, $y$ is the $y$-pixel on the camera and $s$ is the spectral bin. 
All the data preprocessing pipelines were set up with the \texttt{irisreader} Python library\footnote{\url{https://github.com/i4Ds/IRISreader}}.

\subsection{Sampling and cross-validation}

Because the number of available observations is very limited and does most certainly not capture the full variability in the dynamics leading to a flare, splitting of the data into a training and validation set must be handled with care. To get a more robust estimate for the validation set error, we chose to divide the set of observation time windows into three cross-validation folds, stratified by flare class (no flare, M, X). Thereby we took care that no information can leak from the training to the validation set by making sure that spatially and temporally close observation time windows are restricted to the same fold (see observation groups indicated in tables \ref{tab:arobs} and \ref{tab:pfobs}).

The selected observation time windows come at a various number of steps (see tables \ref{tab:arobs} and \ref{tab:pfobs}). We created a dataset that is balanced towards the classes and the groups of observations therein. To this end, the dataset is composed of 10'000 bags, equally split into the classes AR and PF. The bags for each class are then equally divided among the available observation groups and among the groups again equally distributed over the available observation time windows. If the number of bags to sample was smaller or equal to the number of available timesteps in an observation, we subsampled without replacement, otherwise we sampled with replacement and took oversampling from a few observations into account.

\section{Method\label{sec:method}}

Previous works have asserted that Mg II h\&k profiles can serve as a diagnostic for the physics in the solar chromosphere and can be related to different parameters of the solar atmosphere such as temperatures and velocities at different heights \citep{vernazza1981, leenaarts2013_1, leenaarts2013_2, leenaarts2013_3}. In addition, it has been shown that particular types of profiles only appear around flares \citep{panos2018, panoskleint2020}. This leads us to the assumption that individual spectra can be treated as fundamental \textit{instances} with predictive power for flare prediction with IRIS data. However, individual spectra are not labeled, and class labels (AR / PF) only exist on the level of the individual timesteps. Each timestep is a \textit{bag} of a few hundred spectra along the instrument slit. Here, Multiple Instance Learning (MIL) offers a natural choice for investigating flare prediction, and, in particular, for identifying groups of spectra that are key to the prediction process.

Before entering MIL more deeply, let us first clarify the connection of a few important MIL notions to our dataset:

\begin{itemize}
    \item An \textit{instance} is a single Mg II h\&k intensity profile belonging to a pixel on the spectrograph's slit.
    \item A \textit{bag} is the collection of all Mg II h\&k intensity profiles along the spectrograph's slit measured for one particular timestep (exposure) of an observation.
    \item A model predicts at the \textit{instance-level}, when it predicts a quantity on each individual intensity profile, such as an instance label or an attention value.
    \item A model predicts at the \textit{bag-level}, when it predicts a quantity on a single timestep (exposure) of the full spectrograph's slit, such as a bag label.
\end{itemize}

\subsection{Multiple Instance Learning (MIL)\label{sec:mil}}

In a standard classification setting, individual instances are thought to belong to either a positive or a negative class. In a MIL setting, this information is considered latent and hidden from us. Labels are only available at the \textit{bag-level}, but not at the \textit{instance-level}. In the so-called \textit{standard MIL assumption} \citep{fouldsfrank2010}, bags are labeled as negative if they only contain negative instances and positive, if they contain at least one positive instance. This particular (asymmetric) way of propagating instance-level information to the bag-level puts a lot of emphasis on positively labeled instances.

\begin{figure}[ht]
  \begin{center}
  \includegraphics[width=0.8\linewidth]{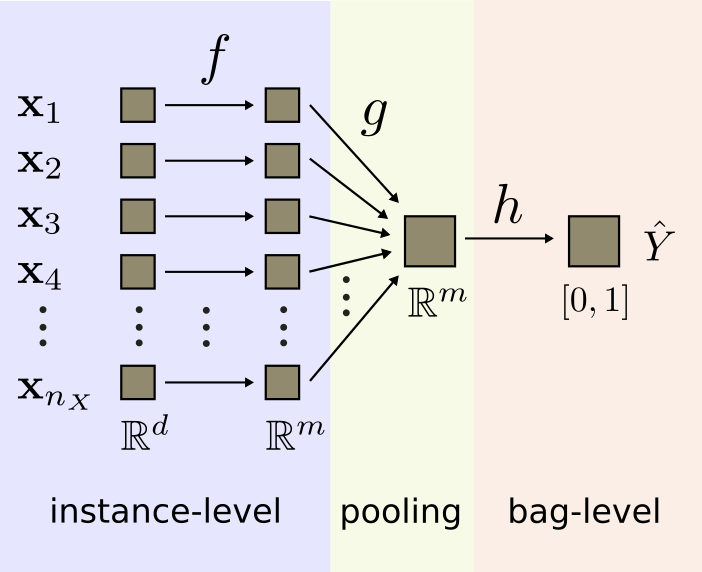}  
  \end{center}
  \caption{Typical setup of a MIL model: A bag $X$ of instances $\left\{\mathbf{x}_i \in \mathbb{R}^d\right\}_{i=1,\cdots, n_X}$ is individually mapped by an instance-level function $f: \mathbb{R}^d \to \mathbb{R}^m$, aggregated to a single bag-level representation by a pooling function $g: \mathbb{R}^m \times \dots \times \mathbb{R}^m \to \mathbb{R}^m$ and converted to a probability by a bag-level classifier $h: \mathbb{R}^m \to [0,1]$. $f$, $g$ and $h$ can be represented by neural networks.}
  \label{fig:mil}
\end{figure}

To deal with a wider variety of problems, the standard MIL assumption can be relaxed by adopting e.g. a \textit{presence-based}, a \textit{threshold-based} or a \textit{collective} assumption (see \citet{fouldsfrank2010} for a review). The threshold-based and collective assumption relax the standard MIL assumption by allowing also a few positive instances in negatively labeled bags, what is often closer to reality.

In the following, we denote quantities at the bag-level with capital letters and quantities at the instance-level with lowercase letters. We denote a bag of instances as $X = \left\{\mathbf{x}_k\right\}_{k=1,\dots,n_X}$ where $\mathbf{x}_k \in \mathbb{R}^d$ is the representation of a single instance and $n_X$ is the number of instances in bag $X$. Individual spectral profiles can be represented as vectors by assigning the i-th interpolated spectral bin to the i-th component of $\mathbf{x}_k$, resulting in vectors with $d{=}240$ dimensions. We attach a label $Y \in \left\{0,1\right\}$ to each bag $X$ and choose $0$ to encode non-flaring active regions (AR) and $1$ to encode pre-flare active regions (PF).

MIL models are usually expressed with a \textit{scoring function} $S$ \citep{abmil} which models the conditional probability for the bag $X$ belonging to the positive class,

\begin{equation}
P(Y=1|X) = S(X), 
\end{equation}
with

\begin{equation}
\hat{Y} = \left\{ \begin{array}{ll}1 & S(X) \geq t\\0 & S(X) < t\end{array}\right.,
\end{equation}
where $t \in (0,1)$ is a threshold value. We decompose $S$ into a function $f$ that operates on the instances $x_k$ of the bag $X$, a permutation-invariant pooling function $g$ that maps from instance to bag-level and a classifier $h$ that operates on the bag-level (Fig.~\ref{fig:mil}), hence 

\begin{equation}
S = h_\psi \circ g_\varphi \circ f_\theta.
\end{equation}

The components $h_\psi$, $g_\varphi$, $f_\theta$ are represented by neural networks parameterized by weights $\theta$, $\varphi$ and $\psi$, that can be trained by minimizing a loss $L\left(Y, P(Y=1|X)\right)$. The decomposition of $S$ is at the heart of this MIL formulation, as it acknowledges that the bag label is an aggregate of the information residing in its different instances.

MIL approaches are usually subdivided into two main branches \citep{fouldsfrank2010}: \textit{instance-based} and \textit{embedding-based}. Below we select and analyze two models, one of each branch.

\subsubsection{Instance-based model}

By design, instance-based models provide the ability to propagate bag labels down to the hidden instance labels. This helps to increase the interpretability of the decisions taken by the model and allows to isolate instances that are responsible for a positive bag label. Prediction at instance-level requires that $f$ acts as a classifier and assigns a probability $p_k \in [0,1]$ to each instance by $p_k = f_\theta(\mathbf{x}_k)$ for belonging to the positive class. The inductive bias is chosen that all learnable parameters are in $f_\theta$, while none are placed in $g$ or $h$. This effectively restricts $g$ to be a pooling function free of learnable parameters and $h$ to be the identity function. The function $g$ is chosen in accordance with the MIL assumption to be adopted: here one could choose $g({p_k}) = \max_k {p_k}$ to satisfy the standard MIL assumption (\textit{max pooling}, one single spectral profile in a bag can already indicate a flare) or $g({p_k}) = \frac{1}{n_X} \sum_k p_k$ as a simple example that satisfies the collective assumption (\textit{average pooling}, a flare requires a few strong or many weaker flare-indicating spectral profiles). While the first approach is hard to train \citep{pinheiro2015} and does not reflect reality, the second does not account for the fact that a few particular spectra might actually be enough to indicate a flare and that a majority of active spectra cannot replace a few very active spectra. As the answer should lie somewhere between the mean and the maximum, we used a smooth interpolation between them with a free parameter $r$ \citep{pinheiro2015, ramon2000}:

\begin{equation}
\label{eq:instance-based}
S_r(X) = g(\{f_\theta(\mathbf{x}_k)\}) = \frac1r \log\left( \frac{1}{n_X} \sum_k e^{r \, f_\theta(\mathbf{x}_k)} \right),
\end{equation}
where $r$ interpolates between the mean ($r\to0$) and the maximum ($r\to\infty$) and can be considered as a tuneable hyperparameter that selects the number of highest probability instances to be considered for classifying the bag. This function is often called '\textit{Log-Sum-Exp}' in the literature (e.g. in \citet{pinheiro2015}), however we find '\textit{Log-Mean-Exp}' more adequate. 

\subsubsection{Embedding-based model}

Embedding-based approaches give up the requirement of predicting at the instance-level and instead try to achieve the best possible performance at the bag-level, without restricting $f$ to return scalar probabilities. To this end, the function $f$ is used to map the individual instances $\mathbf{x}_k$ to an embedding $\mathbf{e}_k \in \mathbb{R}^m$ that is most suitable for bag classification:

\begin{equation}
    \mathbf{e}_k = f_\theta(\mathbf{x}_k).
\end{equation}
The instance-level embeddings are then aggregated to a single bag embedding $\mathbf{E} = g_\varphi(\left\{\mathbf{e}_k\right\})$ that is then used as input for a bag-level classifier network $h_\psi$ with 

\begin{equation}
\label{eq:attention-based}
S(X) = h_\psi(\mathbf{E}) = h_\psi(g_\phi(\{f_\theta(\mathbf{x}_k)\})).
\end{equation}
For this work we chose the attention-based pooling proposed by \citet{abmil} that works under the \textit{weighted} collective assumption \citep{fouldsfrank2010}:

\begin{equation}
\label{eq:attentionpooling}
g_\phi(\{\mathbf{e}_k\}) = \sum_{k=1}^{n_X} a_k \, \mathbf{e}_k,
\end{equation}
where the weights $a_k$ with $\sum_{k=1}^{n_X} a_k = 1$ are computed by a gated attention mechanism

\begin{equation}
\label{eq:attentiondef}
a_{k} = \frac{\exp\left[\mathbf{w}^T \, \left(\tanh\left(\mathbf{V} \mathbf{e}_{k}\right) \odot \sigma\left(\mathbf{U} \mathbf{e}_{k}\right) \right) \right]}{\sum_{j=1}^{n_X} \exp\left[\mathbf{w}^T \, \left(\tanh\left(\mathbf{V} \mathbf{e}_{j}\right) \odot \sigma\left(\mathbf{U} \mathbf{e}_{j}\right)\right)\right] },
\end{equation}
with learned weights $\mathbf{U}, \mathbf{V} \in \mathbb{R}^{l \times m}$ and $\mathbf{w} \in \mathbb{R}^l$ that can be summarized as a set of pooling parameters $\phi = \{\mathbf{U}, \mathbf{V}, \mathbf{w}\}$. 
Because the tanh-activated network alone is not sufficiently non-linear to model complex relationships among instances, it is additionally gated by a sigmoid activation function $\sigma(\cdot)$ \citep{abmil}. The free parameter $l$ controls the complexity of the attention mechanism. In addition, we introduce a sharpening parameter $\gamma$ to control the skewness of the $a_k$ by setting $a_k \mapsto a_k^\gamma / (\sum_{j=1}^{n_X} a_j^\gamma)$, introducing the denominator to ensure that the $a_k$ still sum to one.

The model has the freedom to assign below-average
($<1/n_X$) or above-average ($>1/n_X$) attention values to individual spectral profiles.
For a later comparison of the learned attention values between spectral profiles across different observations (with different $n_X$), the attention values can be brought to a common, normalized scale by defining $\tilde{a}_k = n_X \, a_k$. Then, average attention equals to 1, below-average attention to values $<1$ and above-average attention to values $>1$.

While with embedding-based models we give up the possibility of interpretability at the instance-level, the use of an attention mechanism gives us back some interpretability in form of the per-instance weights in the pooling function.

\subsection{Architecture choices for $f_\theta$, $g_\phi$ and $h_\psi$\label{sec:architecture}}

\begin{figure}[t!]
  \begin{center}
  \includegraphics[width=\linewidth]{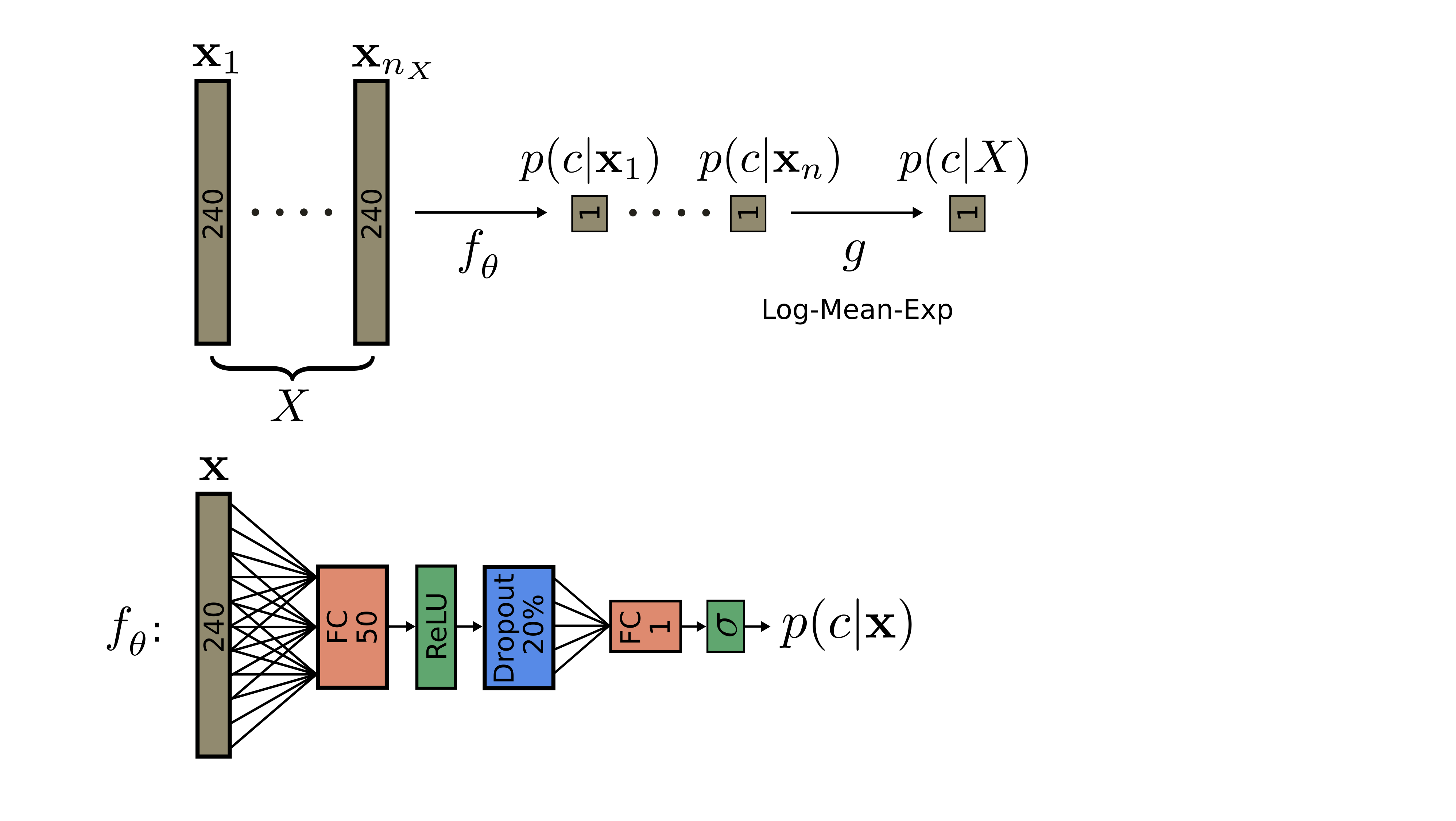}  

  \end{center}
  \caption{Chosen ibMIL model architecture.}
  \label{fig:chosen_ibmil_model}
\end{figure}

\begin{figure}[t!]
  \begin{center}
  \includegraphics[width=\linewidth]{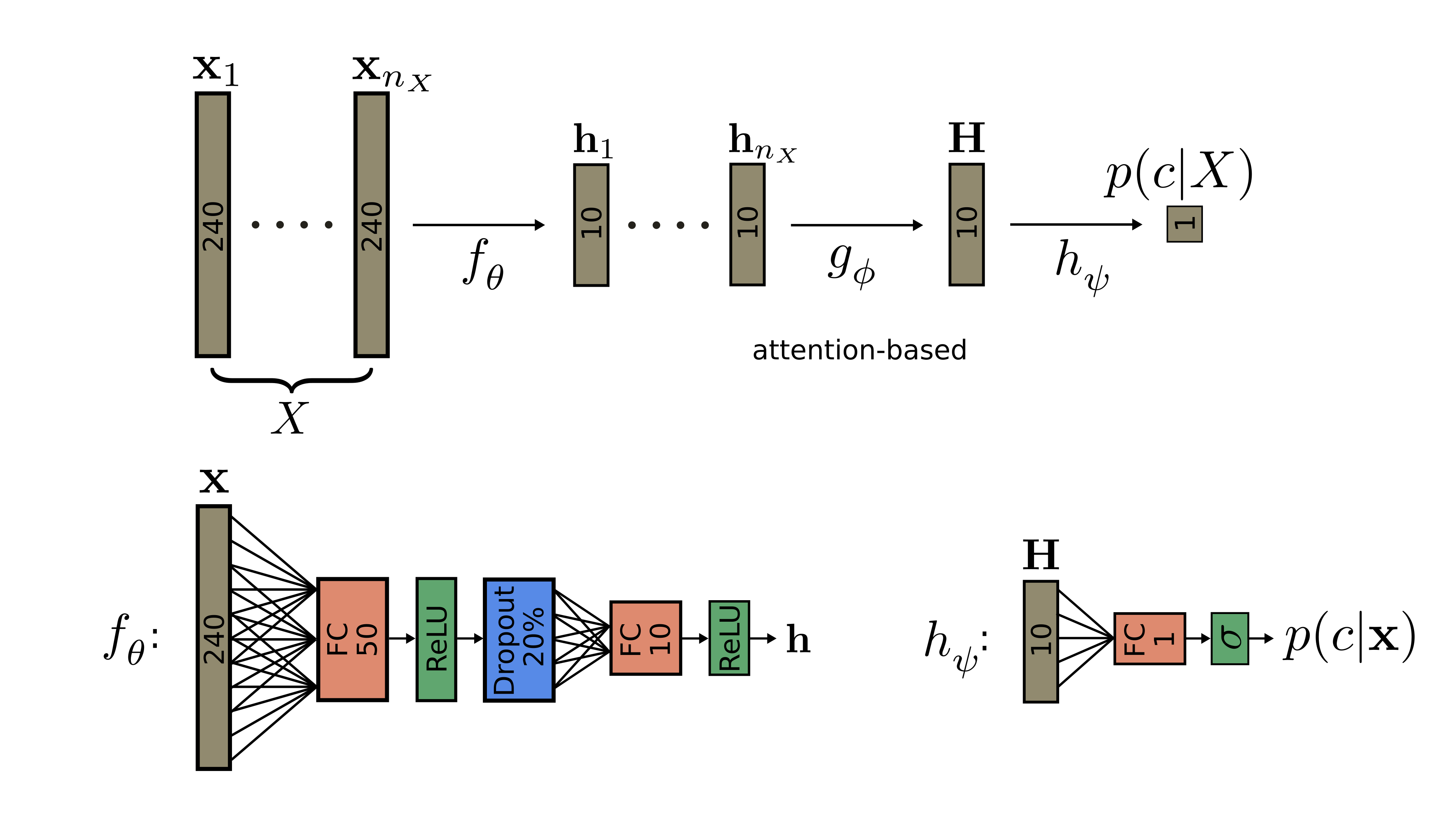}  
  \end{center}
  \caption{Chosen abMIL model architecture.}
  \label{fig:chosen_abmil_model}
\end{figure}

Given the limited dataset with the limited number of observations, we chose rather simple architectures that are nevertheless complex enough to potentially overfit on the training data. Besides the use of a dropout layer \citep{srivastava2014}, we only applied early stopping on the validation loss as a regularization measure. We employed both the instance-based (further called \textit{ibMIL}) and the attention-based (further called \textit{abMIL}) MIL model as introduced in \ref{sec:mil}. The ibMIL model includes $r$ as model-specific hyperparameter. The abMIL model involves the hyperparameters $l$ (complexity of attention mechanism), $m$ (dimension of embedding) and $\gamma$. In this work, we chose $l=10$ and $m=10$. From previous, unpublished experiences with autoencoders, we believe that a latent space with $m=10$ dimensions is sufficiently high-dimensional to represent IRIS spectra (this is supported by \cite{sadykov2021} who use a 4-dimensional space). The choice of $l=10$ is somewhat more arbitrary, however firstly the abMIL models perform already extremely well with $l=10$ (see section \ref{sec:resultsbaglevel}) and secondly we prevent the models from overfitting with early stopping (section \ref{sec:modeltraining}). The particular choice of $l$ and $m$ could definitely be better justified with ablation experiments, however we leave this for later work as it does not change the qualitative results. The role of the parameters $r$ and $\gamma$ will be discussed later.

Our simple model architectures consist only of fully-connected layers, as the most important spectral features can be extracted purely from counts at individual wavelengths and do not require the consideration of neighborhood relations. Nevertheless, convolutional features both in the spectral and spatial domain could potentially improve the predictions and will be explored in future work. For both models we used a 50 nodes fully-connected (FC) hidden layer with ReLU activation in $f_\theta$ followed by 20\% of dropout. For ibMIL we then applied a single fully-connected node with sigmoid activation, while for the abMIL model we applied another ReLU-activated layer with $m$ nodes to map input instances to the $m$-dimensional embedding. The chosen model architectures are sketched in Figs.~\ref{fig:chosen_ibmil_model} (ibMIL) and \ref{fig:chosen_abmil_model} (abMIL).

\subsection{Dealing with heterogeneous bag sizes}

\begin{figure}[h!]
  \begin{center}
  \includegraphics[width=\linewidth]{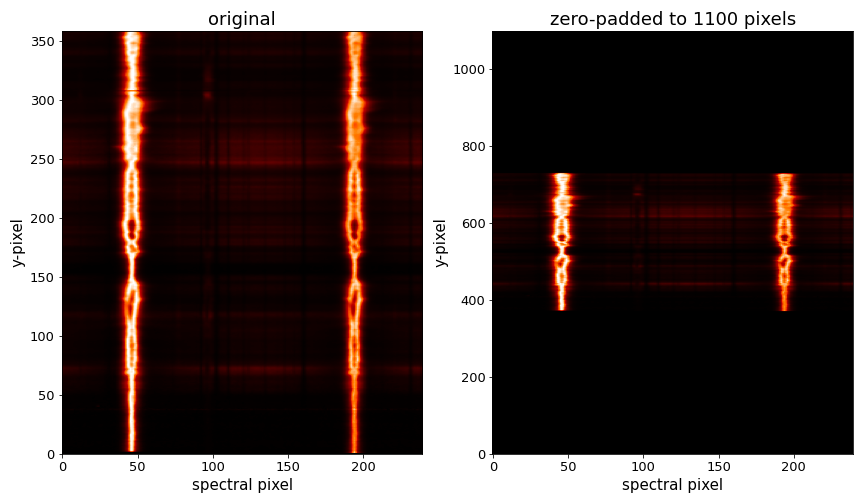}  
  \end{center}
  \caption{Sample timestep of our dataset in its original form (left) and after zero-padding to 1100 y-pixels (right).}
  \label{fig:zeropadding}
\end{figure}

Because the amount of spectral pixels on the $y$-axis per timestep is only constant per observation but varies between different observations, our models need to deal with various bag sizes. At the theoretical level this is not a problem, as the pooling function $g$ does not care about the number of entities it has to aggregate. However deep learning frameworks such as Keras make it much easier to deal with layers of constant input dimension. To this end we zero-pad all the timesteps in our sample to 1100 pixels and store the location of the original data as binary masks (see Fig.~\ref{fig:zeropadding}). For implementation details we refer to appendix \ref{app:masking}.

\subsubsection{Model training\label{sec:modeltraining}}
Instead of a binary cross-entropy loss, we chose a mean absolute error (MAE) loss

\begin{equation}
L(S(X),Y) = \frac1n \sum_i |S(X_i)-Y_i|.
\end{equation}
We found that optimizing MAE loss leads to more stable training. MAE is generally known to perform better in settings with noisy labels \citep{karimi2020, song2022} and we expect that this helps to better deal with outliers that our model cannot identify correctly.

All models were optimized with Adam with parameters $\beta_1=0.9$ and $\beta_2=0.999$. With a learning rate of $0.0001$ we observed stable and efficient learning, for higher learning rates, the training started to oscillate and did not converge. Moreover, more stable training was reached when the input spectra were in addition centered around a mean of $0$ and scaled to standard deviation $1$. We trained each model with a batch size of 32, stopped the training as soon as the validation loss did not improve anymore for at least 20 epochs and stored the model parameters with the best validation loss.

The pre-processed dataset and the implemented models can be found on Github\footnote{\url{https://github.com/chuwyler/irismil}}.

\begin{figure*}[t]
  \centering
  \begin{minipage}{.5\textwidth}
    \centering
    \includegraphics[width=0.9\linewidth]{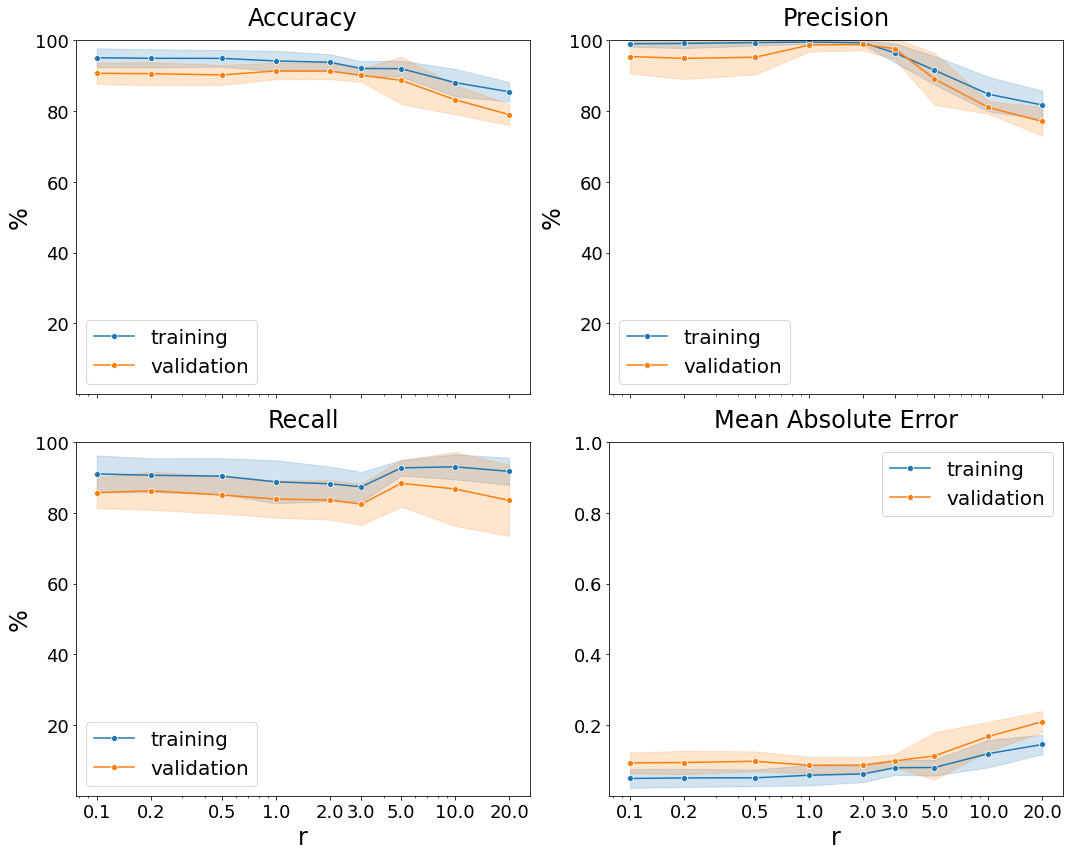}  
  \end{minipage}%
  \begin{minipage}{.5\textwidth}
    \centering
    \includegraphics[width=0.9\linewidth]{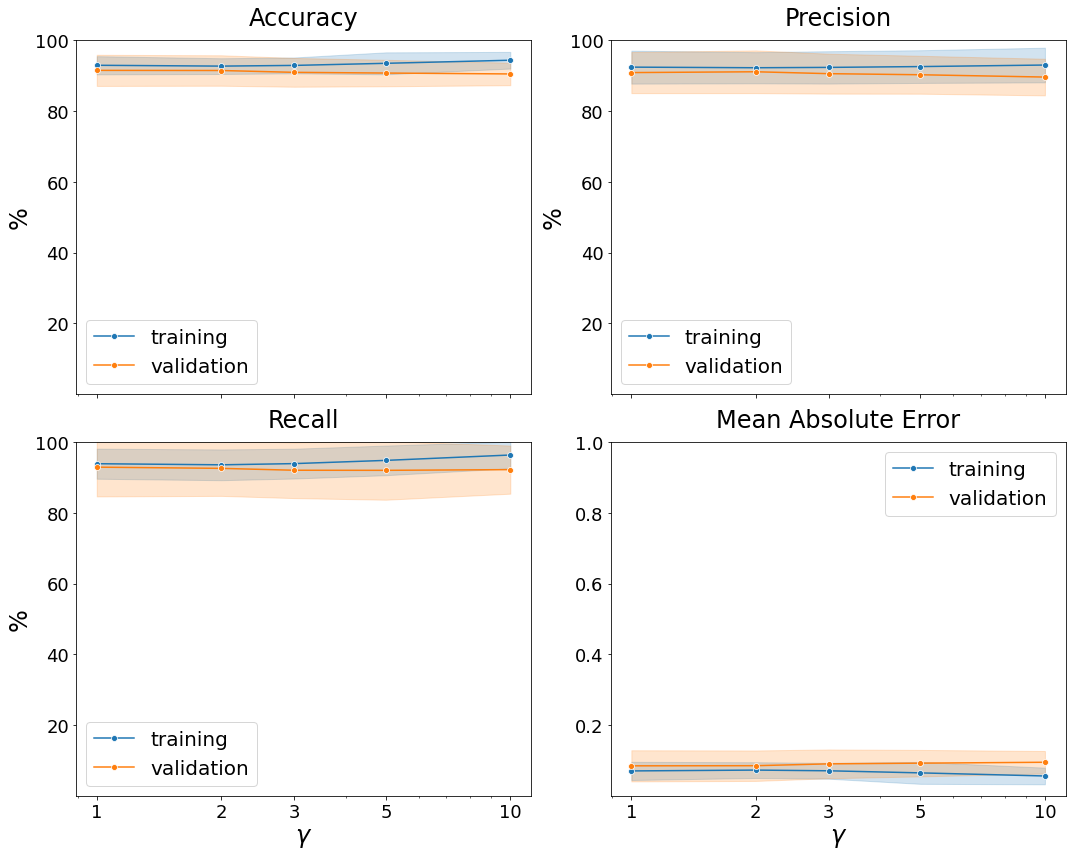}  
  \end{minipage}
  \caption{Distributions of ibMIL (left) and abMIL (right) bag performance measures for different choices of the hyperparameters $r$ or $\gamma$. Both models perform extremely well with accuracies $\gtrsim 90\%$. While the performance of the ibMIL model is sensitive to the choice of $r$ and $r$ seems to have an impact on the precision-recall trade-off, the performance of the abMIL model seems not to be affected by the choice of $\gamma$.}
  \label{fig:bag_performance}
\end{figure*}

\section{Results\label{sec:results}}

In the following we present the results and outputs of the models described in section \ref{sec:method}.
We divide this section into two parts: \textit{bag-level} and \textit{instance-level} prediction. In the bag-level part we evaluate how well the two selected proof-of-concept models perform at the bag-level in predicting a flare within the next 25 minutes. In the instance-level part we discuss the core result of this work: how the predictions at instance-level can serve to increase the explainability of flare prediction models. In particular, we extract groups of spectra that are of special interest to the models and recover similar profiles as in previous works. 

\subsection{Prediction at bag-level\label{sec:resultsbaglevel}}

We trained and evaluated the models described in section \ref{sec:architecture} on the three cross-validation folds described in section \ref{sec:dataset} for different choices of the hyperparameters $r$ or $\gamma$, respectively. In particular, we chose $r \in [1,2,3,5,10,20]$ or $\gamma \in [1,2,3,5,10]$ and trained $10$ models with different parameter initializations for each fold and choice of hyperparameter. The resulting values for accuracy, precision, recall (using a threshold of $0.5$) and mean absolute error (MAE) averaged over all three folds are indicated in Tables \ref{table:ibmil} (ibMIL) and \ref{table:abmil} (abMIL) and visualized in Fig.~\ref{fig:bag_performance}. Both models performed exceptionally well and show a peak validation accuracy on the dataset at $\gtrsim 90\%$ for the smallest chosen values of $r$ or $\gamma$.

For the instance-based model, the validation accuracy drops from around $90\%$ to around $80\%$ when going from $r=1$ (closer to average pooling) to $r=20$ (closer to max pooling). The drop is associated with a loss in precision, which implies that more false positives are returned by the model for larger values of $r$. This makes sense, since higher $r$ bring the model closer to maximum pooling where fewer spectra are considered for the bag-level decision. Here, we hypothesize that it is much more likely that a non-flaring active region has a few flare-like spectra than that a pre-flare observation has only non-flare-like spectra. Interestingly, the attention-based model shows almost constant accuracy, precision and recall for different values of $\gamma$. We assume that the attention-based model is far less susceptible to changes in $\gamma$ because it manages to adapt the learned embeddings due to its higher flexibility. Consequently, we only report instance-level results for $\gamma=1$.

To better understand the behavior of our models, we report the distribution of predicted bag-level probabilities per observation in Fig.~\ref{fig:obs_bag_performance} in the appendix. Thereby we considered for each observation only the learned models that had this particular observation in the validation fold. Probably due to its increased flexibility, the abMIL model distributes bag probabilities much more distinctively than the ibMIL model. For both models, there are clear examples of false positives and false negatives, depending on the value chosen for the parameter $r$ or $\gamma$. 
As can be seen in Fig.~\ref{fig:obs_bag_performance}, most of the false positives and false negatives seem to originate from a few observations, but not all. Clearly, some timesteps of PF observations may not contain information that would announce a flare and some timesteps from AR observations may contain information that would indicate a precursor signal for a flare (we suspect that this might be the reason why the MAE loss leads to a more stable training). 

\begin{figure}[t!]
  \begin{center}
  \includegraphics[width=0.98\linewidth]{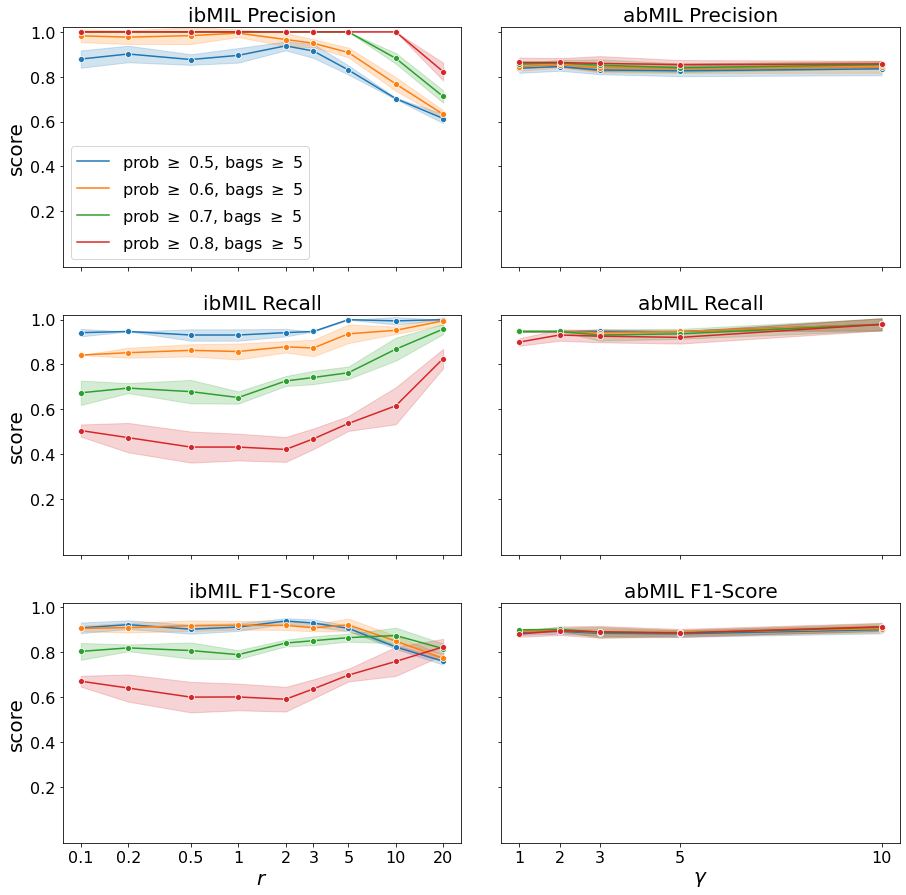}  
  \end{center}
  \caption{Mean precision, recall, F1-score and standard deviations on the observation-level for different values of the parameters $r$ or $\gamma$ and for different detection thresholds for ibMIL (left column) and abMIL (right column) models.}
  \label{fig:obs_level_performance}
\end{figure}

The results given in Fig.~\ref{fig:bag_performance} are an aggregate of the 10'000 bags in our dataset derived from the 37 observations in tables \ref{tab:arobs} and \ref{tab:pfobs}. Since there is a considerable amount of correlation among the different bags, we estimate the performance of our models on the \textit{observation-level} by first aggregating the bag results over the individual observations and then aggregating over all observations. In particular, we do this by specifying detection thresholds through 1) the required probability to count a bag as a pre-flare sign (\textit{bag probability threshold}) and 2) the required number of bags with pre-flare signs to label an observation as pre-flare (\textit{flare bag number}). In Fig.~\ref{fig:obs_level_performance} we show the model performance at observation-level in terms of precision, recall and F1-score over the validation folds for different choices of bag probability thresholds and a minimum of $5$ flare bags against the different chosen values for $r$ and $\gamma$. The detection statistics seem not very sensitive to the chosen required number of flare bags. 
The results of the ibMIL model are sensitive to the choice of probability threshold and, similar to the performance on the bag-level, higher values of the parameter $r$ correspond to lower precision (more false positives) and higher recall (less false negatives). In contrast, the abMIL model is insensitive to the choice of probability threshold and the parameter $\gamma$ influences the abMIL precision-recall trade-off only mildly. Note that the abMIL precision is lower on the observation-level than on the bag-level. This is because of the larger number of false positives, since the bag-level probabilities of some active regions are distributed very broadly (see Fig.~\ref{fig:obs_bag_performance}). The choice of an optimal decision threshold is not further discussed here due to the small amount of labeled data.

In Fig.~\ref{fig:probability_evolution_saliency_maps} we plot the bag-level probability evolution of two selected full pre-flare observations for different values of $r$ and $\gamma$ together with instance-level saliency maps (introduced in section \ref{sec:inference}). While we make use of only 25-minute long AR and PF slices from selected observations, we apply the trained models to the entire observations with lengths of up to several hours to evaluate how our models extrapolate. Intuitively, we would expect the pre-flare probability to increase when getting closer to a flare. However, this is not observed, sometimes probabilities even drop before a flare (e.g. in the ibMIL prediction for the top observation in Fig.~\ref{fig:probability_evolution_saliency_maps}), we currently do not have an explanation for this effect. Interestingly, some of the abMIL models and some of the ibMIL models seem extremely confident in their predictions already hours before the flare onset, despite the carefully designed validation scheme (see section \ref{sec:discussion} for a discussion).

\subsection{Prediction at instance-level\label{sec:inference}}

The learned instance-level probabilities (ibMIL) and attention values (abMIL) can be used to identify spectral profiles that are of particular importance for the models when making their predictions. This may help to understand why the models make a particular prediction at the bag-level and consequently to improve theoretical models in heliophysics that try to explain and predict the emergence of solar flares. In particular, the recovered instance-level predictions can be used to produce so-called \textit{saliency maps} that provide an excellent quick-look tool to identify regions of interest (along the instrument slit) for flare prediction. To visualize the probabilities or attention values predicted by our models, we align the vertically stacked slit pixel predictions horizontally, producing a plot with a spatial $y$-axis and a temporal $x$-axis. In this way, regions of interest on the instrument slit during the evolution of the flare can be highlighted. Saliency maps for two selected full flare observations are provided in Fig.~\ref{fig:probability_evolution_saliency_maps}. Regions with activities that trigger the models are clearly visible.

\subsubsection{Spectra of particular interest to flare prediction\label{sec:particularspectra}}

To extract groups of spectra with particular importance for flare prediction, we used the well-known k-means clustering algorithm \citep{macqueen1967} as a vector quantization method and assessed the distribution of assigned probabilities and attention values per recovered group via mean and standard deviation. The k-means algorithm has been used before to cluster solar spectra \citep{pietarila2007, sanchez2000, viticchi2011, panos2018, sainzdalda2019, woods2021}. We follow this approach since it is simple and easily reproducible. To determine a suitable choice for the number of clusters $k$, we employed the 'elbow rule' \citep{thorndike1953} by visual inspection of the different within-cluster sum-of-squares distances for different choices of $k$. An elbow could approximately be identified at around $k\sim80$. To make sure that also distinguished groups with a smaller number of members can be captured, we chose to 'over cluster' the data with an intuitive choice of $k=128$. It turns out that $k=128$ is sufficient to prove the point of this paper, however in the future this decision could be better substantiated.

\begin{figure*}[h!]
  \begin{center}
   
  \includegraphics[width=\linewidth]{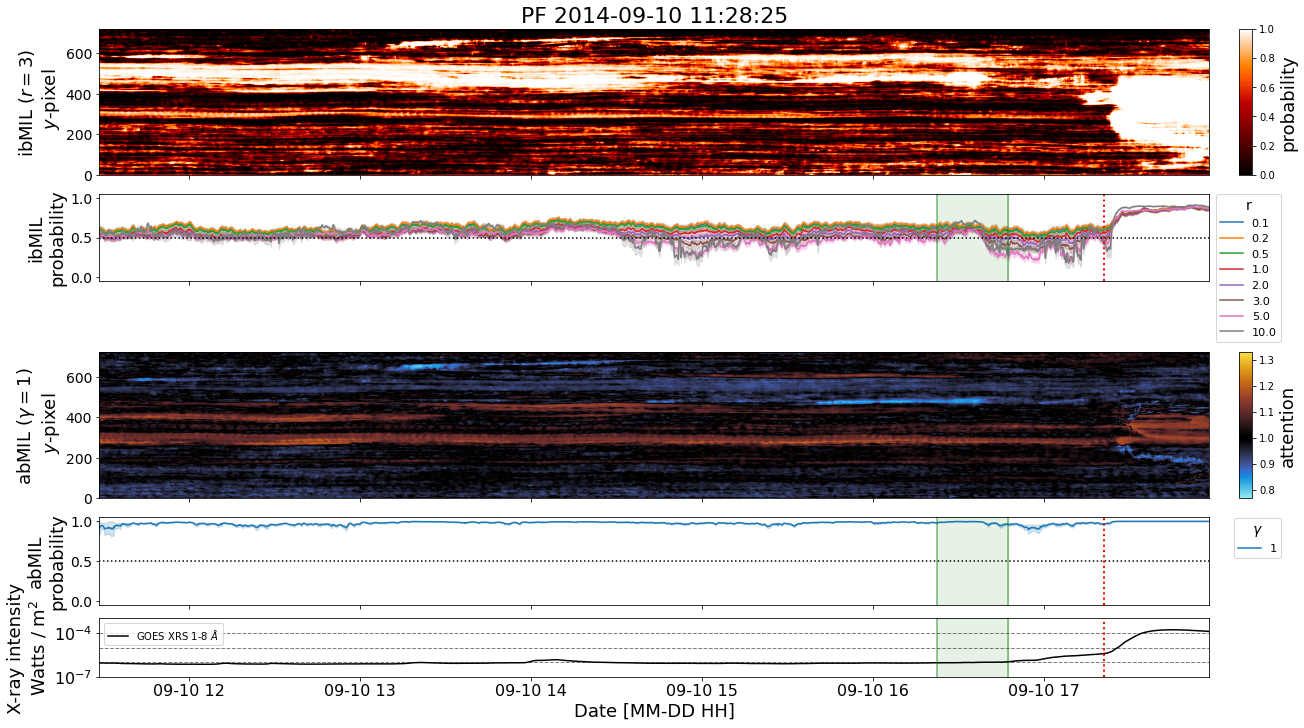} 
  
  \vspace{0.6cm}
  
  \includegraphics[width=\linewidth]{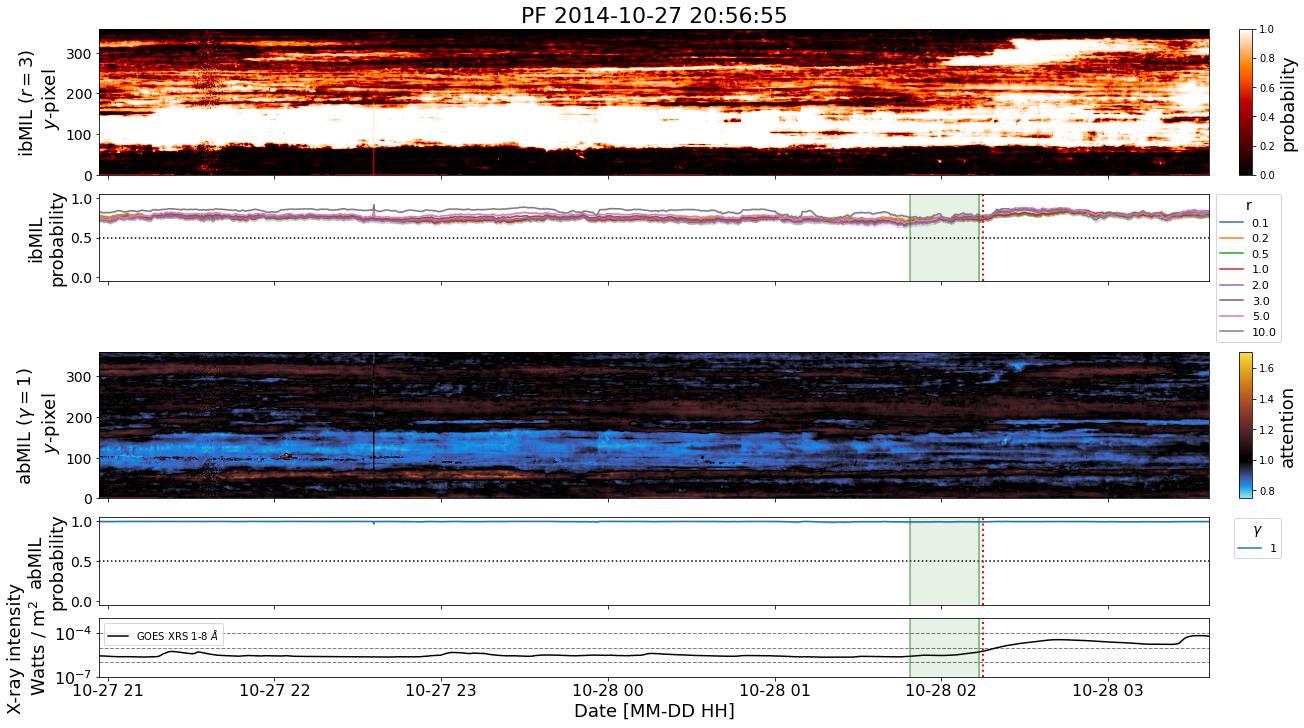}  
  \end{center}
  \caption{Saliency maps and bag-level probability evolutions for two selected full flare observations (\#23 and \#28, indicated by start date in the titles) created by the ibMIL (first and second row) and abMIL (third and fourth row) models that performed best when the particular observation was in the validation set. The saliency maps feature the probabilities recovered by an ibMIL model (with $r=3$) and the normalized attention values recovered by an abMIL model (with $\gamma=1$). The start of the flare is indicated by a red dashed line and the observation time window that entered our dataset is shaded in green. The bottom plot features the X-ray intensity in 1-8 \AA{} measured by NASA's GOES satellite.}
  \label{fig:probability_evolution_saliency_maps}
\end{figure*}

\begin{figure*}[h!]
  \begin{center}
  \includegraphics[width=\linewidth]{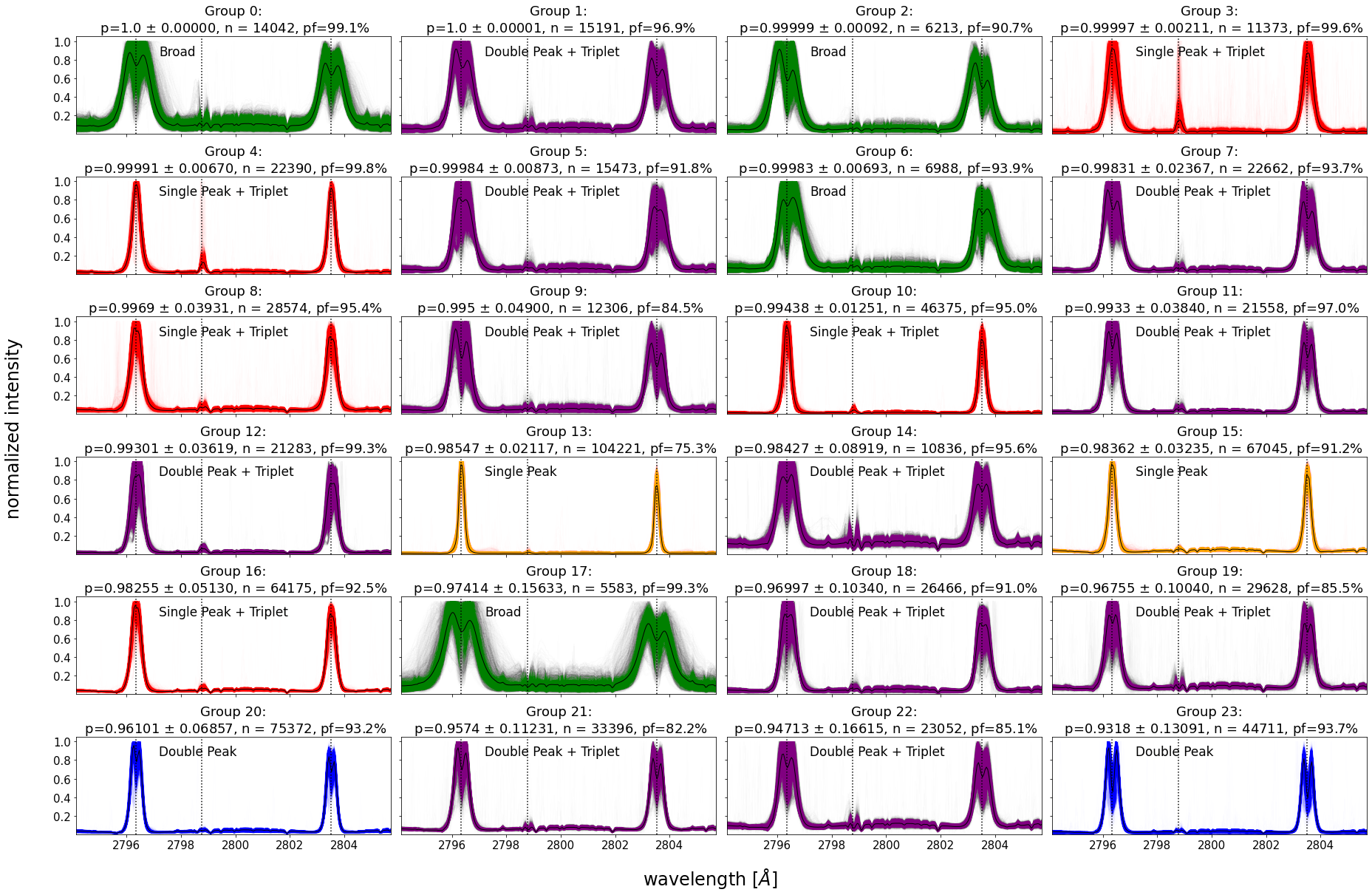}  
  \end{center}
  \caption{Top 24 spectral groups in terms of mean ibMIL probability ($r=1$), sorted by mean probability in descending order. For each group the mean probability (p) and standard deviation ($\pm$), the number of assigned spectra (n) and what percentage of the assigned spectra originates from pre-flare observations (pf) are indicated. In addition to the group centroid (black), a sample of at maximum 10'000 assigned spectra is visualized (in color). The different colors stand for different characteristic overarching groups: broad double-peaked spectra (green), single-peaked spectra with (red) and without (orange) triplet red wing emission and double-peaked spectra with (purple) and without (blue) triplet red wing emission. The dotted vertical lines represent the positions of the k, triplet and h line cores respectively. All the recovered groups with probability $\geq0.7$ can be related to one of the above-mentioned overarching groups.}
  \label{fig:characteristic_groups_ibmil}
\end{figure*}

\begin{figure}[h!]
  \begin{center}
  \includegraphics[width=\linewidth]{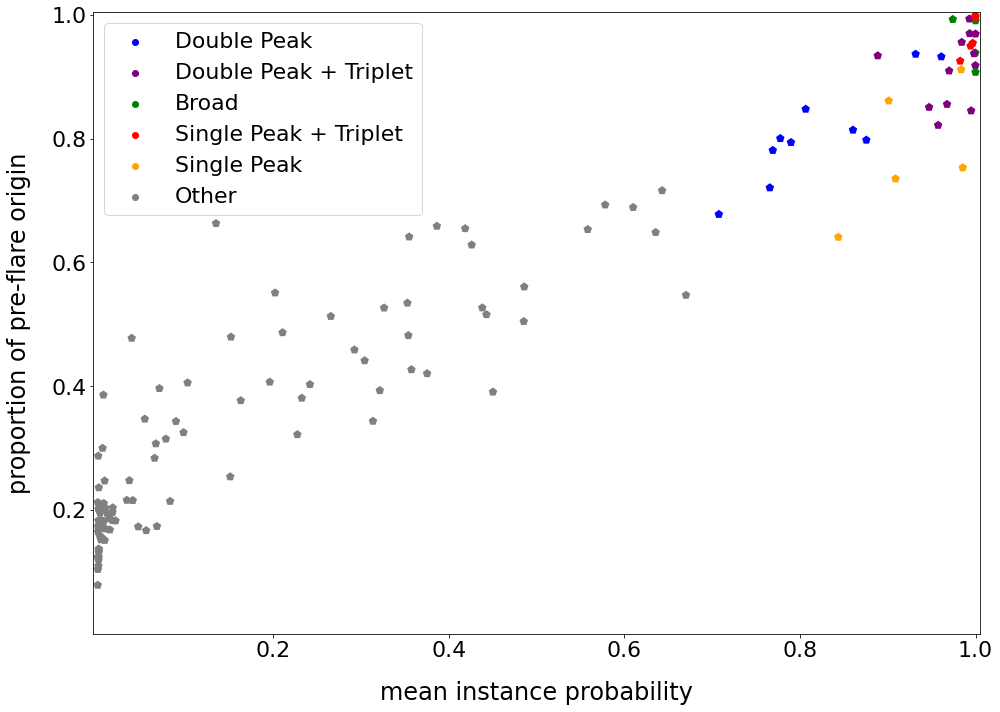}  
  \end{center}
  \caption{Plot of mean instance probability (ibMIL $r=1$) against the proportion of how many spectra originate from pre-flare observations for all the recovered groups. The groups with mean instance probability $\geq 0.7$ are colored similar to Fig.~\ref{fig:characteristic_groups_ibmil}, the rest is denoted as 'Other' and colored in gray. There is a clear and almost linear relationship between mean instance probability and pre-flare proportion.}
  \label{fig:pf_proportions_ibmil}
\end{figure}

For the groups with mean probability $\gtrsim 0.7$ we discovered comparable overarching groups as in a pure k-means study by \citet{woods2021} characterized as follows: single-peaked spectra with and without triplet red wing emission, double-peaked spectra of high intensity with and without triplet red wing emission and broad mostly double-peaked spectra. In Fig.~\ref{fig:characteristic_groups_ibmil}, the top 24 spectral groups with highest mean of the probabilities assigned by our ibMIL model ($r=1$) are visualized, sorted by mean probability in descending order. In addition to mean probability and standard deviation (p), we also indicate the number of spectra assigned to the group (n) and what percentage of the assigned spectra originates from pre-flare observations (pf). Practically all of the spectral profiles in groups with high mean probabilities have predominantly been sampled from pre-flare observations. Similarly, the groups with low mean probabilities are mainly composed of spectral profiles from non-flaring active regions (Fig.~\ref{fig:pf_proportions_ibmil}).

\begin{figure}[h!]
  \begin{center}
  \includegraphics[width=\linewidth]{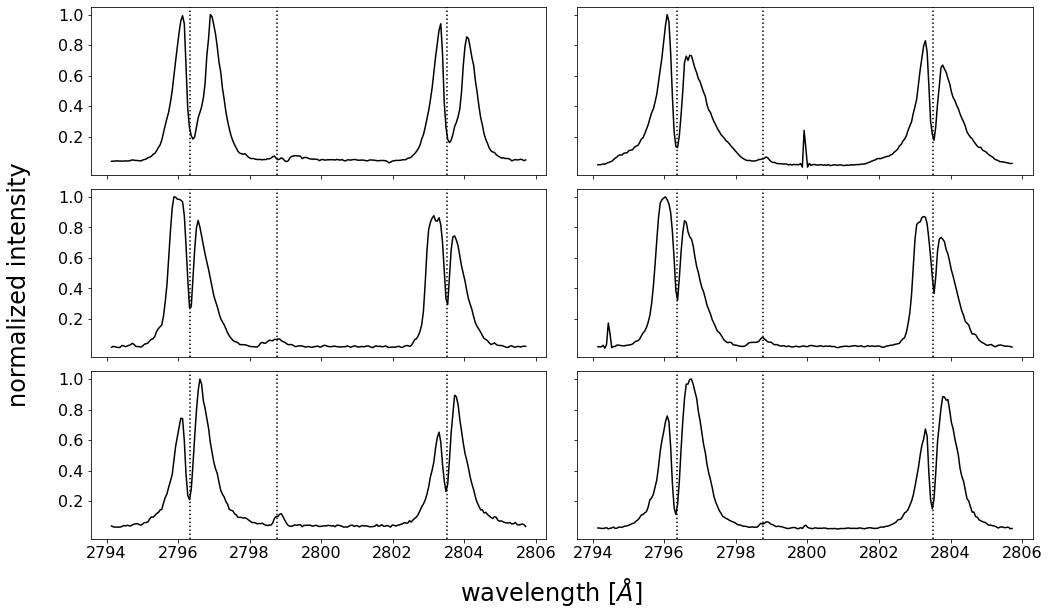}  
  \end{center}
  \caption{Some examples of broad irregular profiles with probability $p=1.0$.}
  \label{fig:broad_irregular_profiles}
\end{figure}

This is the core result of our work. The shallow ibMIL model managed to reproduce state-of-the-art results as a side product of the weakly-supervised prediction problem. As more deeply explored in \citet{woods2021}, the features characterizing the recovered overarching groups can be well related to the physics associated with solar flares:

\begin{itemize}
    \item Emission in the two blended red wing triplets (appearing in red and purple groups) is typically associated with a rapid temperature increase in the lower chromosphere \citep{pereira2015} and is often observed in flares and other high-energy phenomena, in particular also in explosive events accompanied by UV bursts \citep{vissers2015, peter2014, schmit2014}. In some cases (particularly in groups 14, 19, 22), even double-peaked triplet emission can be observed. According to \citet{pereira2015}, these are mostly caused by a temperature distribution along the optical depth with several rapid variations in height rather than just one single dominant increase.
    
    \item Single-peaked Mg II h and k lines (appearing in red and orange groups) have been observed in flares and the umbr\ae{} of sunspots, where the line cores that are usually reversed in quiet-sun conditions (see Fig.~\ref{fig:mgII_example}) go into emission. In contrast to flares, the single-peaked profiles around sunspots are very narrow and do generally not show additional triplet red wing emission. The single-peaked profiles with strong triplet red wing emission from groups 3 and 4 seem to be similar to the universal flaring profile identified by \citet{panos2018} that typically appears over flare ribbons or also small-scale explosive events \citep{panos2018, rubiodacosta2017, zhu2019, kerr2015}. 
    
    \item High peak intensity (corresponding to low continuum in our normalization scheme) is correlated with high chromospheric gas temperatures \citep{kerr2015} and observed in all of the recovered groups.
    
    \item Broad profiles (appearing in green group) are possibly associated with unresolved up and downflows of up to 200 km/s by \citet{rubiodacosta2017}.

\end{itemize}
The high variance in especially the 'broad' and 'double-peaked $+$ triplet' groups suggests  that the k-means algorithm with $k=128$ groups does not provide the perfect granularity for all groups: Division of the double-peaked profiles into further subgroups reveals clusters of various peak ratios corresponding to up and downflows of plasma in the upper chromosphere \citep{leenaarts2013_1} that could be further investigated in future work. In addition, group 17 discloses extremely broad profiles (examples are visualized in Fig.~\ref{fig:broad_irregular_profiles}) with assigned probabilities of practically $1.0$, some with extremely deep central core reversals. These spectra are also discovered by \citet{woods2021} and indicated as \textit{irregular profiles with broad wings}.

The ibMIL model assigns high probabilities to spectral profiles that are highly correlated with an upcoming flare. As such, it rates how important individual spectra are for \textit{only} a positive prediction (what we refer to as 'asymmetric explanation'). In contrast, the abMIL model assigns high attention values to spectral profiles that are of particular importance for the prediction in general, regardless whether it is positive or negative ('symmetric explanation'). Here, accordingly, spectral profiles can act as flare precursors or exactly the opposite --- indicators that clearly no flare is going to take place within the next few hours. As a result, a naive sorting of the clusters by average attention weights does not provide the best flare precursor candidates as obtained in the ibMIL case (as illustrated in Fig.~\ref{fig:characteristic_groups_abmil}). While both the ibMIL and the abMIL model provide explanations for their predictions, the asymmetric explanations by the ibMIL model are better suited for the particular use-case of identifying possible flare precursor candidates. 

\begin{figure*}[htp!]
  \begin{center}
  \includegraphics[width=\linewidth]{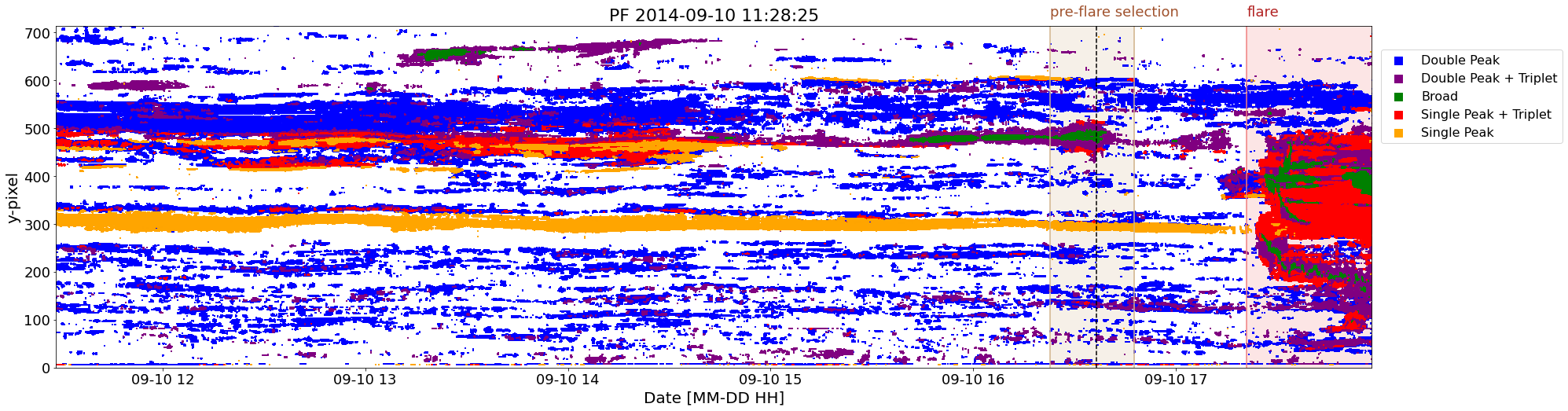}  
  \end{center}
  \caption{Groups recovered from the entire 10 September 2014 flare observation. The duration of the flare and the 25 minutes pre-flare timespan selected for our dataset are highlighted. The flaring part appears to be dominated by single-peaked spectra with triplet red wing emission and broad spectral profiles, however this statement is of limited validity, since the clustering was done only on pre-flare data and does not necessarily extend to spectral profiles observed in flares. In the pre-flare part, high importance is assigned to a mass flow with broad profiles and triplet red wing emission in the middle of the upper part of the slit. This is clearly visible in Fig.~\ref{fig:pf0910_sjioverplot}, a slit-jaw image obtained from the pre-flare selection (indicated by dashed black line). The yellow region (single-peaked without triplet) coincides with the sunspot umbra in Fig.~\ref{fig:pf0910_sjioverplot}.}
  \label{fig:pf0910_groups}
\end{figure*}

\begin{figure}[htp!]
  \begin{center}
  \includegraphics[width=0.8\linewidth]{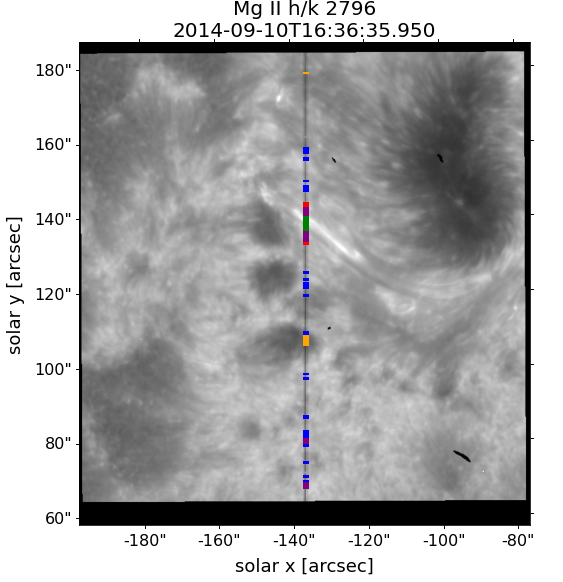}  
  \end{center}
  \caption{Slit-jaw image taken during the pre-flare selection of the 10 September 2014 flare observation. The instrument slit is indicated by a vertical line and the different overarching groups from Figs. \ref{fig:characteristic_groups_ibmil} and \ref{fig:pf0910_groups} are highlighted. The mass-flow in the upper middle part is clearly visible and is captured by spectra from the broad (green) group, accompanied by spectra from the triplet emission groups (red, purple).}
  \label{fig:pf0910_sjioverplot}
\end{figure}

Given the recovered overarching groups, we are able to examine more deeply why the ibMIL model assigned high probabilities to particular pixels on the slit. As an example, we provide a visualization of the recovered groups for the 10 September 2014 flare in Fig.~\ref{fig:pf0910_groups} (the corresponding probabilities are visualized in the upper part of Fig.~\ref{fig:probability_evolution_saliency_maps}). 
We notice that high importance is given to broad spectra (colored in green) surrounded by spectra with triplet red wing emission (colored in red) in the upper part of Fig.~\ref{fig:pf0910_groups} within the pre-flare selection window. In Fig.~\ref{fig:pf0910_sjioverplot}, a slit-jaw image taken during this phase where the instrument slit in the center is overplotted with the corresponding group colors is displayed. It appears that the observed groups of spectra are caused by a larger mass flow in this region. Similar plots are shown for two example non-flaring active regions in the appendix in Fig.~\ref{fig:recovered_ar_groups}.

In Fig.~\ref{fig:obsgroup_overview} in the appendix we list how often the selected overarching groups of average ibMIL probability $\geq 0.7$ appear in the different observation windows included in our dataset. For the pre-flare active region observations typically the frequencies of one or more of the groups are strongly enhanced, presumably the ones that are connected to the underlying processes that led to the flare. We could not identify a single group that is always strongly present in the pre-flare phase, though.

\section{Discussion\label{sec:discussion}}

We have demonstrated that instance-based MIL models are more suited to identify possible flare precursors at the cost of bag-level performance, while embedding-based models show a very high bag-level performance but lack explainability. The missing explainability can partially be restored by analyzing attention weights, however the identification of pre-flare precursors is far less easy than originally expected: a particular spectral profile can be considered as important by the model and be assigned a high attention value, however the model does not suggest a straightforward procedure to associate the spectral profile with a positive or negative bag-level class. 

The parameters $r$ (ibMIL) or $\gamma$ (abMIL) control the skewness of the distribution of assigned instance-level probabilities or attention values, respectively. While high values of $r$ lead the ibMIL model to focus only on the strongest predictors and significantly impact the precision-recall trade-off, the choice of $\gamma$ has an almost negligible impact in case of the abMIL model. As stated previously, we believe that the reason for this is that the abMIL model can flexibly accommodate distortions resulting from modified parameters. We find that the results presented in this paper are qualitatively similar for different choices of $r$ or $\gamma$ and propose to use values that lead to the best bag-level validation set accuracy ($r=1$ and $\gamma=1$ in the case of our dataset).

With the chosen very simple models and evaluation strategy we see significant potential for further improvements. The models neglect the spatial structure (correlations between neighboring spectra) and the dynamics (correlations between successive spectra) --- aspects that both might be crucial for accurate flare prediction and proper identification of flare-preceding patterns. In addition, our models are too shallow to capture more advanced features than just the positions of emission peaks, line widths and the continuum intensity. Deeper models could find more advanced features such as the the peak ratios in or between the k and h lines. The latter has been reported to be reduced before the onset of and during flares \citep{panos2018, panoskleint2020}. Finally, we observed that k-means is not perfectly suited for the analysis of model predictions when working with normalized spectra: Important very localized features such as triplet emission or peak ratios in the k or h line are washed out by the Euclidean distance measure. Also, Doppler shifts and instrument artifacts such as cosmic ray spikes can move two spectra of otherwise similar shape to different, unrelated groups. In this work, we could alleviate these problems partially by choosing significantly more groups ($k=128$) than proposed by the elbow rule ($k=80$). 

However, our models seem nevertheless to be able to discern between non-flaring active regions and pre-flare active regions with very high accuracy on the validation set, the abMIL model in particular. In some examples they predict flares with high confidence even hours before the flare (see e.g.~the top observation in Fig.~\ref{fig:probability_evolution_saliency_maps}). We assume that such a high performance would not be expected in an operational scenario. For direct comparison to other flare prediction approaches, we would need to incorporate prior probabilities for AR and PF events. This is mainly due to the fact that our models were trained and validated on a balanced dataset. Moreover, in contrast to other publications that predict flares within the next 12 or 24 hours, we only consider a time span of approximately half an hour before flare onset (25 minutes plus margin) for the evaluation of our models. Some works suggest that typical pre-flare signatures such as small-scale heating events might be much more common closer to the flare onset \citep{panoskleint2020, woods2021}. 

Finally, finding a truly representative sample of non-flaring and pre-flare active regions is a challenging task in all flare prediction problems, since the distribution of the underlying population is not known well enough a priori. Even though our dataset was created in a joint effort between heliophysics and machine learning experts, there is always the potential danger for a selection bias that could impact the reported model performance estimates.

\section{Conclusion and Outlook}

We demonstrated the application of the Multiple Instance Learning (MIL) paradigm to predict flares given spectral Mg II h\&k data recorded by the IRIS satellite and in particular to extract groups of spectral profiles that act as flare precursor signals. The reasons why we advocate the use of MIL for IRIS data are two-fold: Firstly, MIL offers a way to exploit the available weak labels on the observation-level by providing a model that links the data from individual spectra up to the observation or bag-level in a well-defined way. This allows training the model with weakly-labeled data and subsequent predictions on the observation or bag-level. Secondly, the specifically selected MIL models allow to propagate the weak observation labels down the other way to individual, instance-level spectral profiles (ibMIL) or at least allow the identification of spectral profiles that are of particular importance for the prediction of the model (abMIL).

We evaluated both an instance-based approach (ibMIL) and an embedding-based approach incorporating an attention mechanism (abMIL) on a balanced dataset sampled from 18 non-flaring and 19 pre-flare active region observations. Both models performed reasonably well on the bag-level with accuracies of $\gtrsim 90\%$ in predicting whether or not an active region is going to produce a flare within the next $\sim$25 minutes. The abMIL model performed slightly better than the ibMIL model at the cost of reduced explainability of its results. For both models, saliency maps of the computed probabilities or attention values provide insights into the decision process over the time evolution of the observation. 
For the core result of this paper, we used the instance-level predictions by the ibMIL model to extract a set of typical pre-flare spectral profiles. The recovered groups are consistent with the results of other works in the domain \citep{woods2021} and seem to be related to changes in the temperature, velocity and density distribution induced by small-scale explosive events that have been reported to occur tens of minutes before a flare.

In view of developing a production-ready flare prediction model based on IRIS data, several aspects need to be explored:

Firstly, we plan to further extend our sample of non-flaring active region observations by evaluating our models on as many available IRIS observations as possible and applying an active learning strategy. In addition, we intend to design an appropriate prior to extend the predictions of our models to more realistic class imbalance.
Secondly, we propose to use models with substantially more capacity and suitable regularization to identify more complex pre-flare precursor features, such as asymmetries in the Mg II h and k lines and the Mg II k/h peak ratio.
Thirdly, accurate flare prediction requires the identification of spatio-temporally evolving patterns. We plan to set up a pooling function that takes into account spatial structure through the use of convolutional neural networks. The dynamical aspect of the problem could be included by using an embedding-based approach and pooling the embeddings of the individual timesteps with a recurrent neural network.
Fourthly, the inclusion of data from other spectral windows observed by IRIS, such as C II, Si IV or O IV will almost certainly provide additional and very valuable information (as explored in \citet{panosmultiline1} and \citet{panosmultiline2}).
Fifthly, we plan to extend the tools to analyze our results, since k-means does not seem to be optimal to divide spectral profiles into groups. We plan to evaluate alternative clustering strategies (such as Deep Embedded Clustering \citep{xie2016}), possibly applied directly to the embeddings recovered by the MIL models.

We have shown that MIL provides a very valuable interpretable modeling approach for the prediction of solar flares. It looks very promising to also be applied to other solar data, such as images recorded by the HMI or AIA instruments on the SDO satellite.

\section*{Acknowledgements}

We thank Brandon Panos for valuable comments.  
This research has been supported by SNSF NFP75 grant no. 407540\_167158.
IRIS is a NASA small explorer mission developed and operated by LMSAL with mission operations executed at NASA Ames Research center and major contributions to downlink communications funded by ESA and the Norwegian Space Centre. We would like to extend our gratitude to LMSAL for kindly providing us direct access to their IRIS data storage.

\bibliography{mybibfile}

\begin{appendix}

\section{Using masking to account for heterogeneous bag sizes\label{app:masking}}

Once all the bags are zero-padded to $n=1100$ pixels, the whole dataset can be represented by the data cube $\mathbf{X} \in \mathbb{R}^{b\times n\times d}$ and the vector $\mathbf{Y} \in \mathbb{R}^b$, where $b=10'000$ represents the number of bags, $n=1100$ the number of pixels and $d=240$ the number of spectral bins.
The masks are represented by a matrix $\mathbf{M} \in  \mathbb{R}^{b\times n}$ and the number of (non-zero) instances per bag is

\begin{equation}
n_i = \sum_{k=1}^n M_{ik}.
\end{equation}
The ibMIL instance-level classifier $f_\theta$ learns the matrix of probabilities $\mathbf{P} \in \mathbb{R}^{b \times n}$ given the matrix $\mathbf{X}$ which is then pooled to compute the bag score

\begin{equation}
S(X) = \frac1r \log \left( \frac{1}{n_i} \sum_{k=1}^{n} M_{ik} e^{r \, P_{ik}} \right),
\end{equation}
where the multiplication with $M_{ik}$ makes sure that only masked probabilities enter the bag aggregate.

Similarly, the abMIL approach computes the bag score from a matrix of bag embeddings $\mathbf{E} \in \mathbb{R}^{b\times m}$, where $m$ is the latent dimension of the bag embeddings, that is aggregated as 

\begin{equation}
E_{ij} = \sum_{k=1}^{n} e_{ikj} A_{ik}   ,
\end{equation}
where $\mathbf{e} \in \mathbb{R}^{b\times n\times m}$ represents the individual embeddings computed by $f_\theta$ and $\mathbf{A} \in \mathbb{R}^{b\times n}$ represents the attention values learned by the pooling $g$. These have to be appropriately masked:

\begin{equation}
A_{ik} = \frac{M_{ik} Q_{ik}}{\sum_{j=1}^n M_{ij} Q_{ij}}
\end{equation}
with $Q_{ik} = \exp\left[\mathbf{w}^T \, \tanh\left(\mathbf{V} \, \mathbf{e}_{ik}\right) \odot \sigma\left(\mathbf{U} \,  \mathbf{e}_{ik}\right) \right]^\gamma$ and the vectors $\mathbf{e}_{ik} \in \mathbb{R}^m$.

\begin{figure*}
  \begin{center}
  \includegraphics[width=0.90\linewidth]{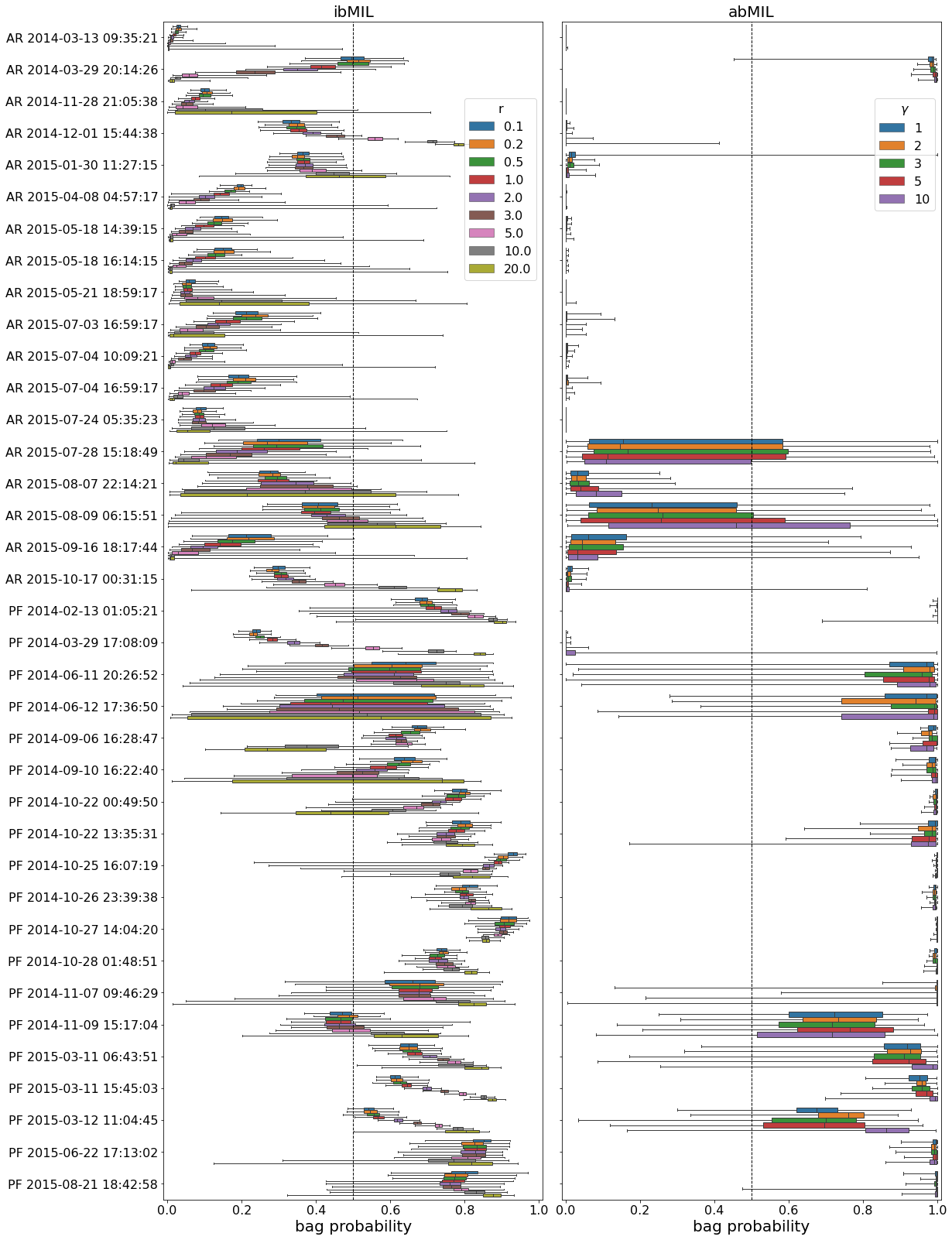}  
  \end{center}
  \caption{Recovered ibMIL and abMIL validation bag probability distributions for individual observations. The non-flaring active region (AR) observations are located in the top half and the pre-flare observations (PF) in the bottom-half. The individual distributions are indicated with a box plot whose whiskers stretch to the minimum and maximum of the distribution. The free parameter $r$ has a particular impact on the distribution of ibMIL bag probabilities, while the impact of $\gamma$ on the abMIL bag probabilities is less pronounced. Note that the two misclassified observations from 29 March 2014 are from different locations on the Sun.
}
  \label{fig:obs_bag_performance}
\end{figure*}

\clearpage

\begin{sidewaystable}
    \centering
    \begin{tabular*}{\linewidth}{cccccccccccc}
    \hline
    \hline
    \# & Start & Duration & Observation Mode & FOV Center & Steps & Cadence & Slit Pixels & Group & Validation Fold \\
    & & (min) & & (arcsec) & & (sec) & & & & \\
    \hline
   0 & 2014-03-13 09:35:21 & 24.9 & Very large sit-and-stare  & (522, 23) & 162 & 9.3 & 548 & 0 & 0 \\
    1 & 2014-03-29 20:14:26 & 24.8 & Medium sit-and-stare  & (686, -166) & 90 & 16.8 & 401 & 1 & 1 \\
    2 & 2014-11-28 21:05:38 & 24.9 & Very large sit-and-stare  & (-34, -322) & 157 & 9.6 & 1096 & 2 & 1 \\
    3 & 2014-12-01 15:44:38 & 24.8 & Large sit-and-stare  & (-79, -329) & 155 & 9.7 & 779 & 3 & 1 \\
    4 & 2015-01-30 11:27:15 & 20.3 & Very large dense 4-step raster  & (-726, 161) & 240 & 20.2 & 548 & 4 & 2 \\
    5 & 2015-04-08 04:57:17 & 24.9 & Very large sit-and-stare  & (46, -118) & 285 & 5.3 & 548 & 5 & 0 \\
    6 & 2015-05-18 14:39:15 & 25.0 & Large coarse 4-step raster  & (298, -98) & 280 & 21.4 & 777 & 6 & 0 \\
    7 & 2015-05-18 16:14:15 & 25.0 & Large coarse 4-step raster  & (313, -95) & 280 & 21.1 & 777 & 6 & 0 \\
    8 & 2015-05-21 18:59:17 & 24.9 & Very large sit-and-stare  & (-386, 398) & 277 & 5.4 & 1095 & 7 & 2 \\
    9 & 2015-07-03 16:59:17 & 24.8 & Large sparse 8-step raster  & (-189, 203) & 272 & 43.8 & 777 & 8 & 1 \\
    10 & 2015-07-04 10:09:21 & 24.9 & Very large sit-and-stare  & (64, 174) & 165 & 9.1 & 548 & 9 & 1 \\
    11 & 2015-07-04 16:59:17 & 24.8 & Large sparse 8-step raster  & (17, 202) & 272 & 43.9 & 777 & 8 & 1 \\
    12 & 2015-07-24 05:35:23 & 24.8 & Large sit-and-stare  & (537, 203) & 163 & 9.2 & 388 & 10 & 2 \\
    13 & 2015-07-28 15:18:49 & 24.9 & Medium coarse 4-step raster  & (-229, -289) & 164 & 36.5 & 207 & 11 & 2 \\
    14 & 2015-08-07 22:14:21 & 24.7 & Large coarse 8-step raster  & (541, 125) & 160 & 74.3 & 777 & 12 & 0 \\
    15 & 2015-08-09 06:15:51 & 25.0 & Large coarse 8-step raster  & (230, -370) & 160 & 75.3 & 777 & 13 & 0 \\
    16 & 2015-09-16 18:17:44 & 24.8 & Medium coarse 16-step raster  & (-578, -346) & 704 & 33.6 & 207 & 14 & 0 \\
    17 & 2015-10-17 00:31:15 & 24.9 & Large sit-and-stare  & (-561, -232) & 478 & 3.1 & 388 & 15 & 2 \\
    \hline
    \end{tabular*}
    \caption{Selected non-flare active regions (AR) in our sample\label{tab:arobs}}
\end{sidewaystable}

\begin{sidewaystable}
    \centering
    \begin{tabular*}{\linewidth}{cccccccccccc}
    \hline
    \hline
    \# & Start & Duration & Observation Mode & FOV Center & Steps & Cadence & Slit Pixels & Group & Validation Fold  \\
       & & (min) & & (arcsec) & & (sec) & & & & \\
    \hline
    18 & 2014-02-13 01:05:21 & 25.1 & Large coarse 8-step raster  & (133, -99) & 280 & 41.7 & 773 & 16 & 1 \\
    19 & 2014-03-29 17:08:09 & 24.8 & Very large coarse 8-step raster  & (507, 282) & 160 & 71.9 & 1094 & 17 & 2 \\
    20 & 2014-06-11 20:26:52 & 25.0 & Medium coarse 8-step raster  & (-777, -305) & 560 & 21.3 & 203 & 18 & 2 \\
    21 & 2014-06-12 17:36:50 & 25.0 & Medium coarse 8-step raster  & (-691, -303) & 560 & 21.4 & 203 & 19 & 0 \\
    22 & 2014-09-06 16:28:47 & 17.8 & Large sit-and-stare  & (-704, -299) & 114 & 9.5 & 775 & 20 & 0 \\
    23 & 2014-09-10 16:22:40 & 24.9 & Large sit-and-stare  & (-121, 125) & 160 & 9.4 & 776 & 21 & 1 \\
    24 & 2014-10-22 00:49:50 & 24.6 & Large sit-and-stare  & (-343, -316) & 91 & 16.4 & 776 & 22 & 1 \\
    25 & 2014-10-22 13:35:31 & 23.8 & Very large coarse 8-step raster  & (-249, -298) & 88 & 131.2 & 1094 & 23 & 1 \\
    26 & 2014-10-25 16:07:19 & 25.0 & Large sit-and-stare  & (406, -319) & 280 & 5.4 & 387 & 24 & 2 \\
    27 & 2014-10-26 23:39:38 & 24.8 & Large sit-and-stare  & (637, -287) & 93 & 16.2 & 387 & 25 & 2 \\
    28 & 2014-10-27 14:04:20 & 6.3 & Large coarse 8-step raster  & (727, -293) & 120 & 25.6 & 387 & 26 & 0 \\
    29 & 2014-10-28 01:48:51 & 24.8 & Large sit-and-stare  & (787, -270) & 93 & 16.2 & 387 & 27 & 0 \\
    30 & 2014-11-07 09:46:29 & 25.1 & Large coarse 16-step raster  & (-659, 224) & 976 & 24.3 & 386 & 28 & 1 \\
    31 & 2014-11-09 15:17:04 & 8.6 & Large coarse 4-step raster  & (-219, 205) & 56 & 37.4 & 775 & 29 & 1 \\
    32 & 2015-03-11 06:43:51 & 24.9 & Large coarse 8-step raster  & (-420, -193) & 160 & 75.0 & 776 & 30 & 0 \\
    33 & 2015-03-11 15:45:03 & 24.8 & Large coarse 4-step raster  & (-352, -198) & 288 & 16.8 & 388 & 30 & 0 \\
    34 & 2015-03-12 11:04:45 & 22.4 & Large sit-and-stare  & (-190, -190) & 258 & 5.2 & 388 & 30 & 0 \\
    35 & 2015-06-22 17:13:02 & 24.8 & Large sparse 16-step raster  & (70, 187) & 704 & 33.1 & 387 & 31 & 2 \\
    36 & 2015-08-21 18:42:58 & 25.2 & Medium dense 32-step raster  & (-452, -337) & 480 & 102.1 & 207 & 32 & 2 \\
    \hline
    \end{tabular*}
    \caption{Selected pre-flare active regions (PF) in our sample\label{tab:pfobs}}
\end{sidewaystable}

\begin{table*}
    \begin{tabular*}{0.45\linewidth}{ccccc}
    \hline
    \hline
    \# & Flare Start & Flare Class  \\
       & (min) & (GOES) \\
    \hline
    18 & 2014-02-13 01:31:59 & M1.8 \\
    19 & 2014-03-29 17:35:37 & X1.0 \\
    20 & 2014-06-11 20:53:00 & M3.9 \\
    21 & 2014-06-12 18:02:58 & M1.0 \\
    22 & 2014-09-06 16:50:04 & M1.1 \\
    23 & 2014-09-10 17:21:02 & X1.6 \\
    24 & 2014-10-22 01:16:06 & M8.7 \\
    25 & 2014-10-22 14:01:43 & X1.6 \\
    26 & 2014-10-25 16:41:18 & X1.0 \\
    27 & 2014-10-27 00:06:06 & M1.3 \\
    28 & 2014-10-27 14:11:59 & X2.0 \\
    29 & 2014-10-28 02:15:03 & M3.4 \\
    30 & 2014-11-07 10:12:49 & M1.0 \\
    31 & 2014-11-09 15:27:04 & M2.3 \\
    32 & 2015-03-11 07:10:10 & M1.8 \\
    33 & 2015-03-11 16:11:00 & X2.1 \\
    34 & 2015-03-12 11:38:18 & M1.4 \\
    35 & 2015-06-22 17:39:00 & M6.5 \\
    36 & 2015-08-21 19:09:55 & M1.1 \\
    \hline
    \end{tabular*}
    \caption{Flare start times and magnitudes for the pre-flare active regions in our sample}
    \label{tab:flare_start_times}
\end{table*}

\begin{table*}
\begin{tabular}{lllllll}
\toprule
 \textbf{$r$} &  \textbf{train acc [\%]} &   \textbf{val acc [\%]} & \textbf{train prec [\%]} &  \textbf{val prec [\%]} &  \textbf{train rec [\%]} &    \textbf{val rec [\%]} \\
\midrule
        0.1 &  95.1 $\pm$ 2.7 &  90.7 $\pm$ 2.9 &  99.1 $\pm$ 0.9 &  95.4 $\pm$ 4.8 &  91.1 $\pm$ 5.2 &   85.7 $\pm$ 4.3 \\
        0.2 &  94.9 $\pm$ 2.6 &  90.6 $\pm$ 3.3 &  99.1 $\pm$ 1.3 &  94.9 $\pm$ 5.8 &  90.7 $\pm$ 4.8 &   86.3 $\pm$ 5.4 \\
        0.5 &  94.9 $\pm$ 2.4 &  90.2 $\pm$ 2.8 &  99.4 $\pm$ 0.9 &  95.2 $\pm$ 4.9 &  90.4 $\pm$ 5.1 &   85.1 $\pm$ 5.3 \\
        1.0 &  94.2 $\pm$ 2.9 &  91.4 $\pm$ 2.4 &  99.5 $\pm$ 0.5 &  98.7 $\pm$ 1.9 &  88.8 $\pm$ 6.1 &   83.9 $\pm$ 5.3 \\
        2.0 &  93.8 $\pm$ 2.3 &  91.3 $\pm$ 2.3 &  99.3 $\pm$ 0.6 &  98.9 $\pm$ 1.6 &  88.3 $\pm$ 4.9 &   83.7 $\pm$ 5.6 \\
        3.0 &  92.1 $\pm$ 2.1 &  90.1 $\pm$ 1.8 &  96.5 $\pm$ 2.7 &  97.7 $\pm$ 3.0 &  87.4 $\pm$ 4.2 &   82.5 $\pm$ 5.8 \\
        5.0 &  92.0 $\pm$ 2.3 &  88.7 $\pm$ 6.7 &  91.6 $\pm$ 4.0 &  89.1 $\pm$ 7.3 &  92.7 $\pm$ 2.2 &   88.4 $\pm$ 6.6 \\
       10.0 &  88.1 $\pm$ 3.9 &  83.2 $\pm$ 4.1 &  84.8 $\pm$ 4.9 &  81.0 $\pm$ 1.8 &  93.1 $\pm$ 3.5 &  86.8 $\pm$ 10.5 \\
       20.0 &  85.5 $\pm$ 2.7 &  79.0 $\pm$ 3.0 &  81.7 $\pm$ 4.0 &  77.1 $\pm$ 4.0 &  91.8 $\pm$ 3.8 &  83.6 $\pm$ 10.1 \\
\bottomrule
\end{tabular}
\caption{ibMIL model results for training and validation accuracy (acc), precision (prec) and recall (rec) for different choices of $r$.\label{table:ibmil}}
\end{table*}

\begin{table*}
\begin{tabular}{lllllll}
\toprule
 \textbf{$\gamma$} &  \textbf{train acc [\%]} &   \textbf{val acc [\%]} & \textbf{train prec [\%]} &  \textbf{val prec [\%]} &  \textbf{train rec [\%]} &   \textbf{val rec [\%]} \\
\midrule
              1 &  93.0 $\pm$ 2.6 &  91.5 $\pm$ 4.4 &  92.4 $\pm$ 4.7 &  90.9 $\pm$ 5.8 &  93.9 $\pm$ 4.3 &  93.0 $\pm$ 8.3 \\
              2 &  92.7 $\pm$ 2.2 &  91.5 $\pm$ 4.3 &  92.3 $\pm$ 4.4 &  91.1 $\pm$ 6.0 &  93.6 $\pm$ 4.4 &  92.6 $\pm$ 7.8 \\
              3 &  92.9 $\pm$ 2.2 &  91.0 $\pm$ 4.1 &  92.4 $\pm$ 4.6 &  90.6 $\pm$ 5.6 &  94.0 $\pm$ 4.2 &  92.1 $\pm$ 7.9 \\
              5 &  93.5 $\pm$ 3.1 &  90.8 $\pm$ 3.7 &  92.6 $\pm$ 4.6 &  90.3 $\pm$ 5.4 &  94.9 $\pm$ 4.2 &  92.1 $\pm$ 8.3 \\
             10 &  94.4 $\pm$ 2.4 &  90.5 $\pm$ 3.2 &  93.0 $\pm$ 4.9 &  89.6 $\pm$ 5.2 &  96.4 $\pm$ 3.8 &  92.3 $\pm$ 6.8 \\
\bottomrule
\end{tabular}
\caption{abMIL model results for training and validation accuracy (acc), precision (prec) and recall (rec) for different choices of $\gamma$.\label{table:abmil}}
\end{table*}

\begin{figure*}[h!]
  \begin{center}
   
  \includegraphics[width=\linewidth]{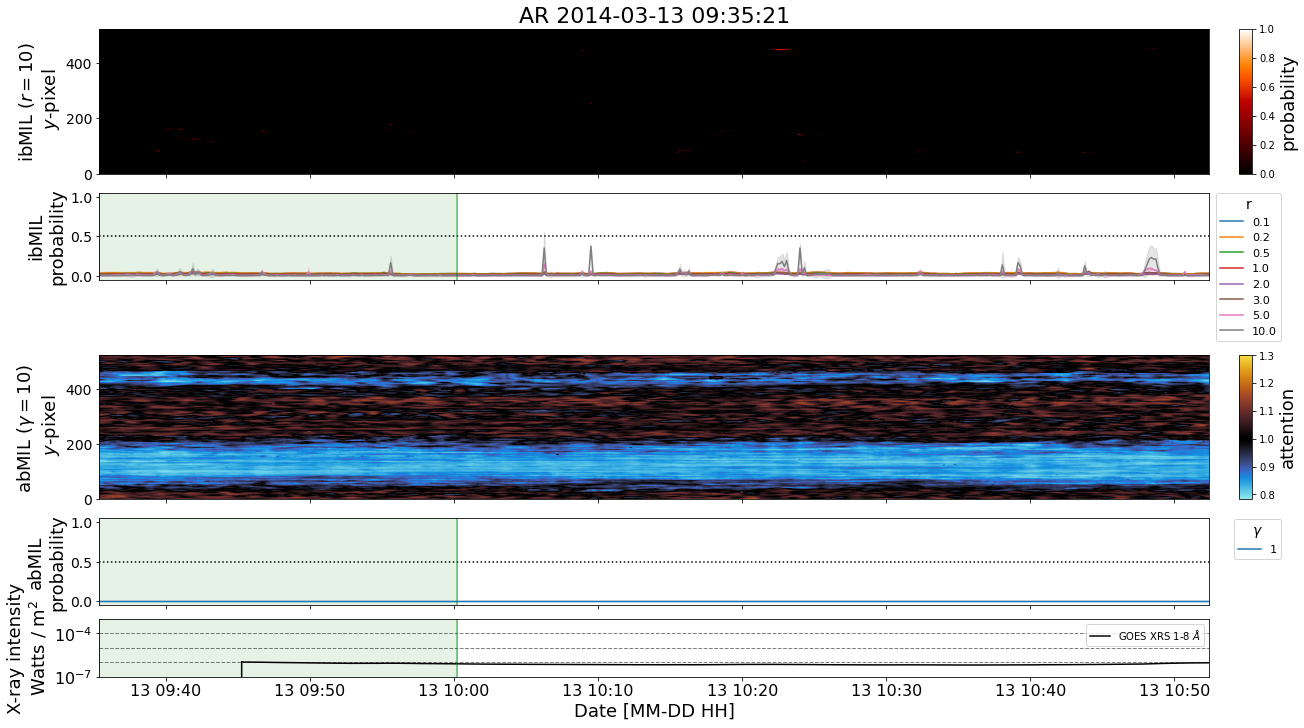} 
  
  \vspace{0.6cm}
  
  \includegraphics[width=\linewidth]{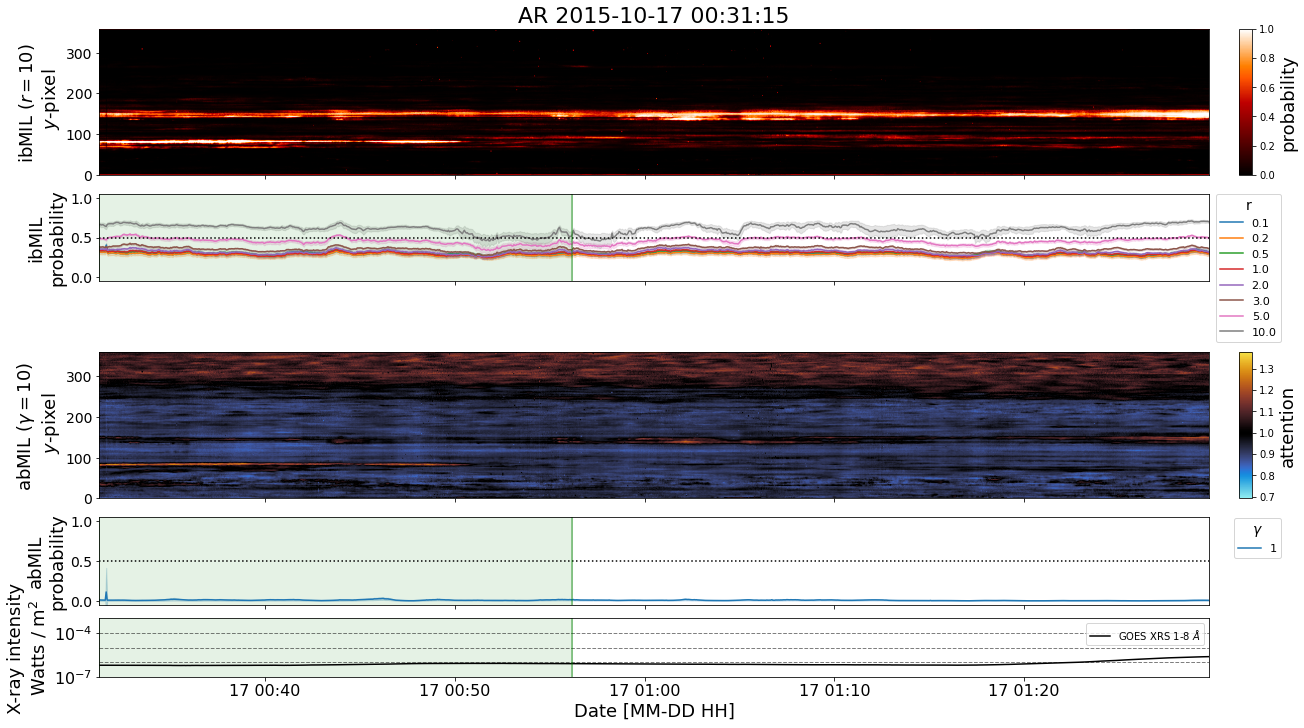}  
  \end{center}
  \caption{Saliency maps and probability evolutions for two selected active region observations (\#0 and \#17) created by the ibMIL (first and second row) and abMIL (third and fourth row) models in a similar fashion as in Fig.~\ref{fig:probability_evolution_saliency_maps}. All the models predict the top observation as a clear negative. In contrast, ibMIL models with $r \gtrsim 5$ falsely predict the bottom observation to be pre-flare because of local brightenings observed around slit pixels 85 and 150. As previously explained, the saliency maps with the attention values are difficult to read.}
  \label{fig:probability_evolution_saliency_maps_ar}
\end{figure*}

\begin{figure*}[h!]
  \begin{center}
   
  \includegraphics[width=\linewidth]{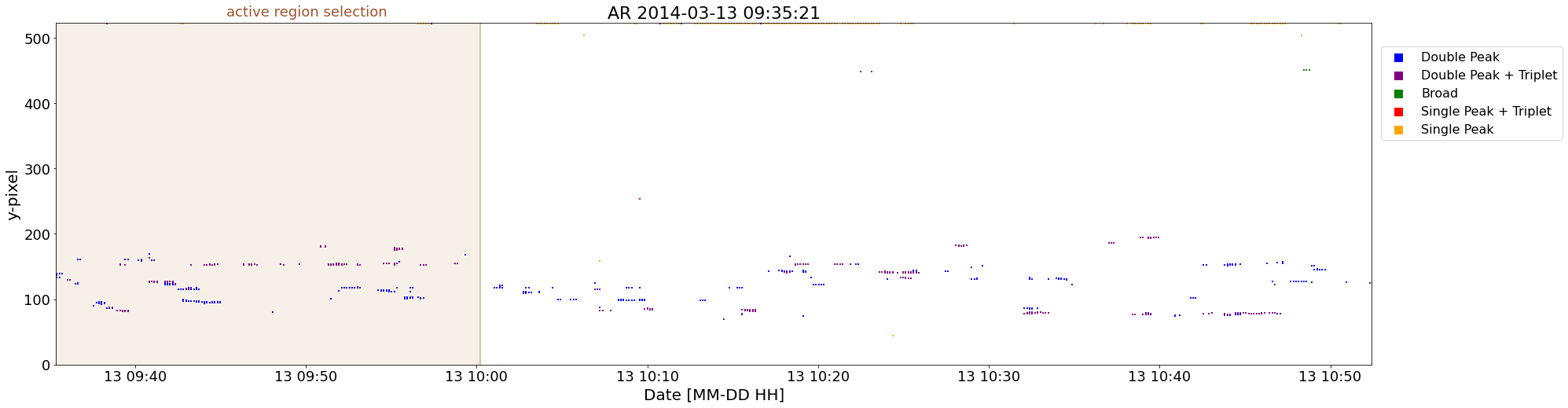} 
  
  \vspace{0.6cm}
  
  \includegraphics[width=\linewidth]{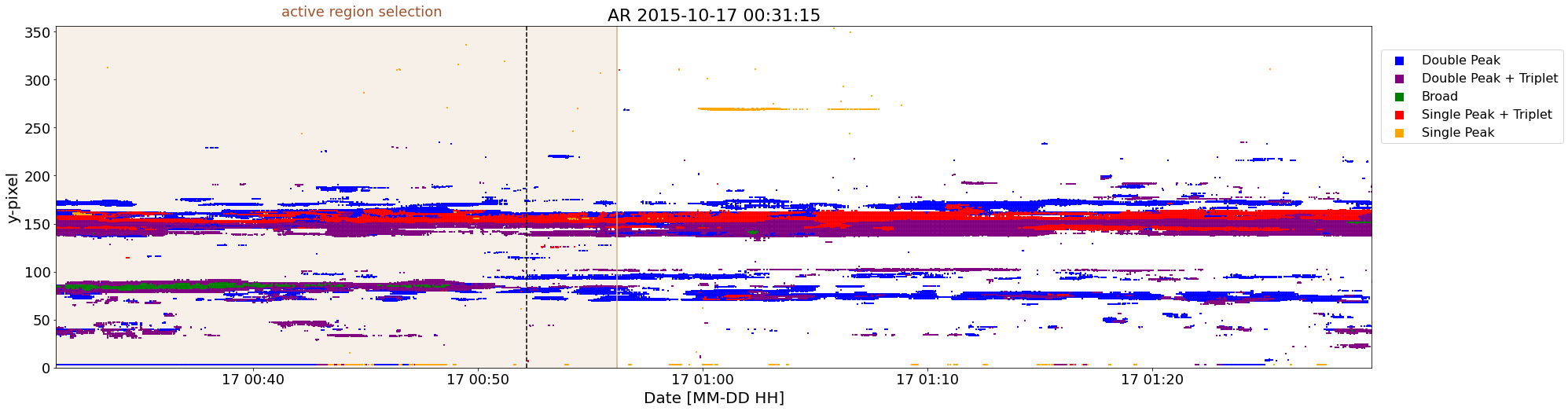}  
  
  \includegraphics[width=0.4\linewidth]{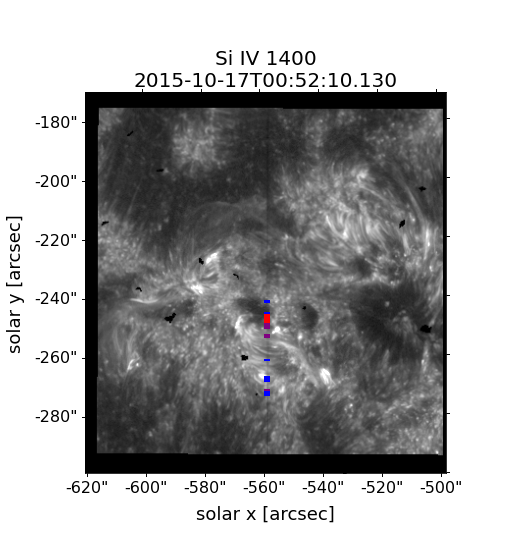}
  
  \end{center}
  \caption{Groups recovered from the entire 03 March 2014 (top) and 17 October 2015 (middle) active region observations, visualized in a similar fashion as in Fig.~\ref{fig:pf0910_groups}. The timespan of 25 minutes selected for our dataset is highlighted ('active region selection'). Practically none of the pre-flare groups are identified in the top observation. The second observation shows more pre-flare-like activity around pixels 150 and 85 (as can be compared with Fig.~\ref{fig:probability_evolution_saliency_maps_ar}). Similar to the September 10 flare (Figs.~\ref{fig:pf0910_groups} and \ref{fig:pf0910_sjioverplot}), this might be due to mass flows and associated brightenings around these pixels, as illustrated in the slit-jaw image overplot in the bottom figure (taken at the position of the dashed black line in the middle plot).}
  \label{fig:recovered_ar_groups}
\end{figure*}

\begin{sidewaysfigure*}[t]
  \begin{center}
  \includegraphics[width=\linewidth]{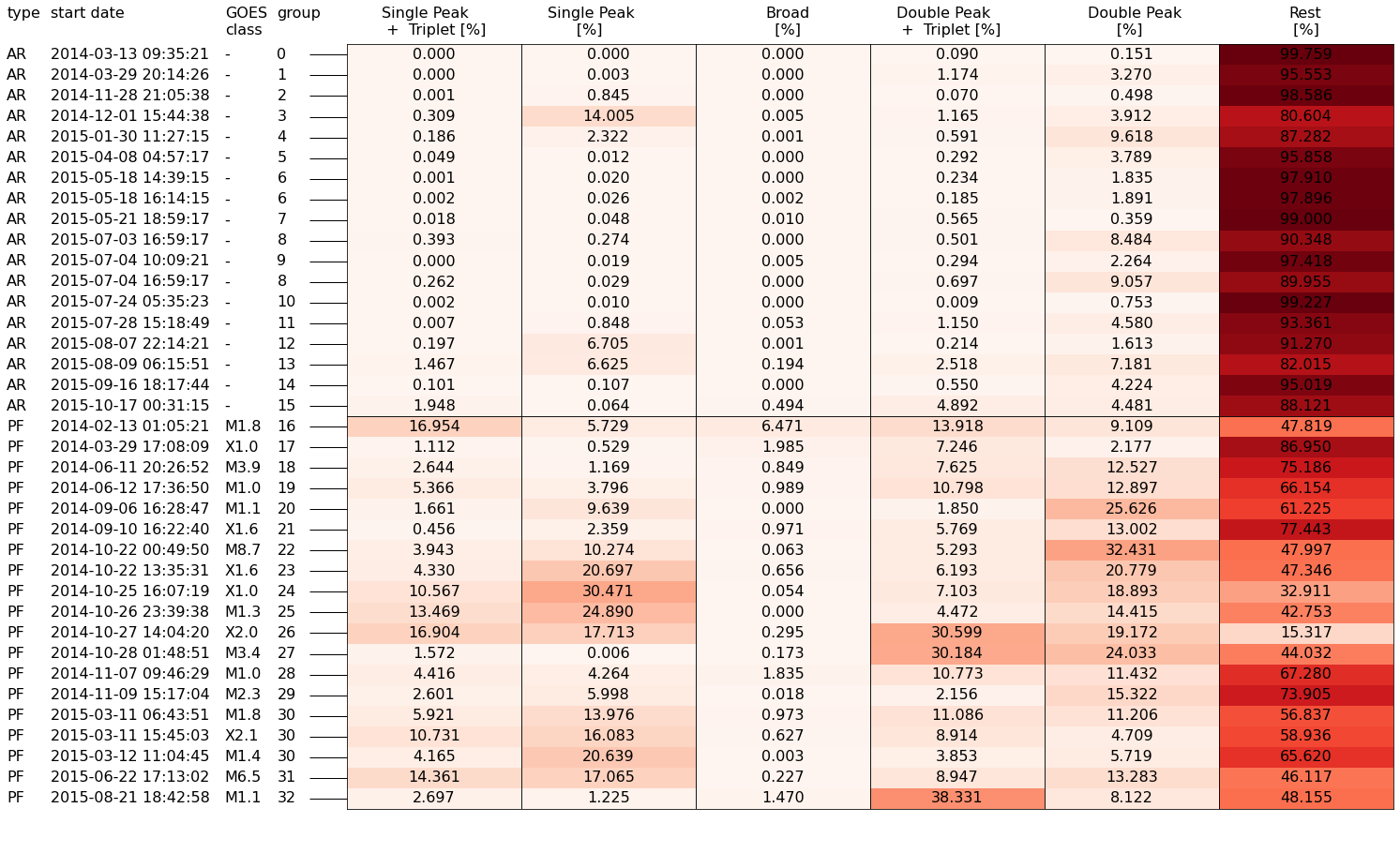}  
  \end{center}
  \caption{Frequency distributions of the recovered overarching groups observed in the different observation windows included in our dataset. The frequencies of one or more of the selected groups are strongly enhanced in pre-flare active region observations.}
  \label{fig:obsgroup_overview}
\end{sidewaysfigure*}

\begin{figure*}[h!]
  \begin{center}
  \includegraphics[width=\linewidth]{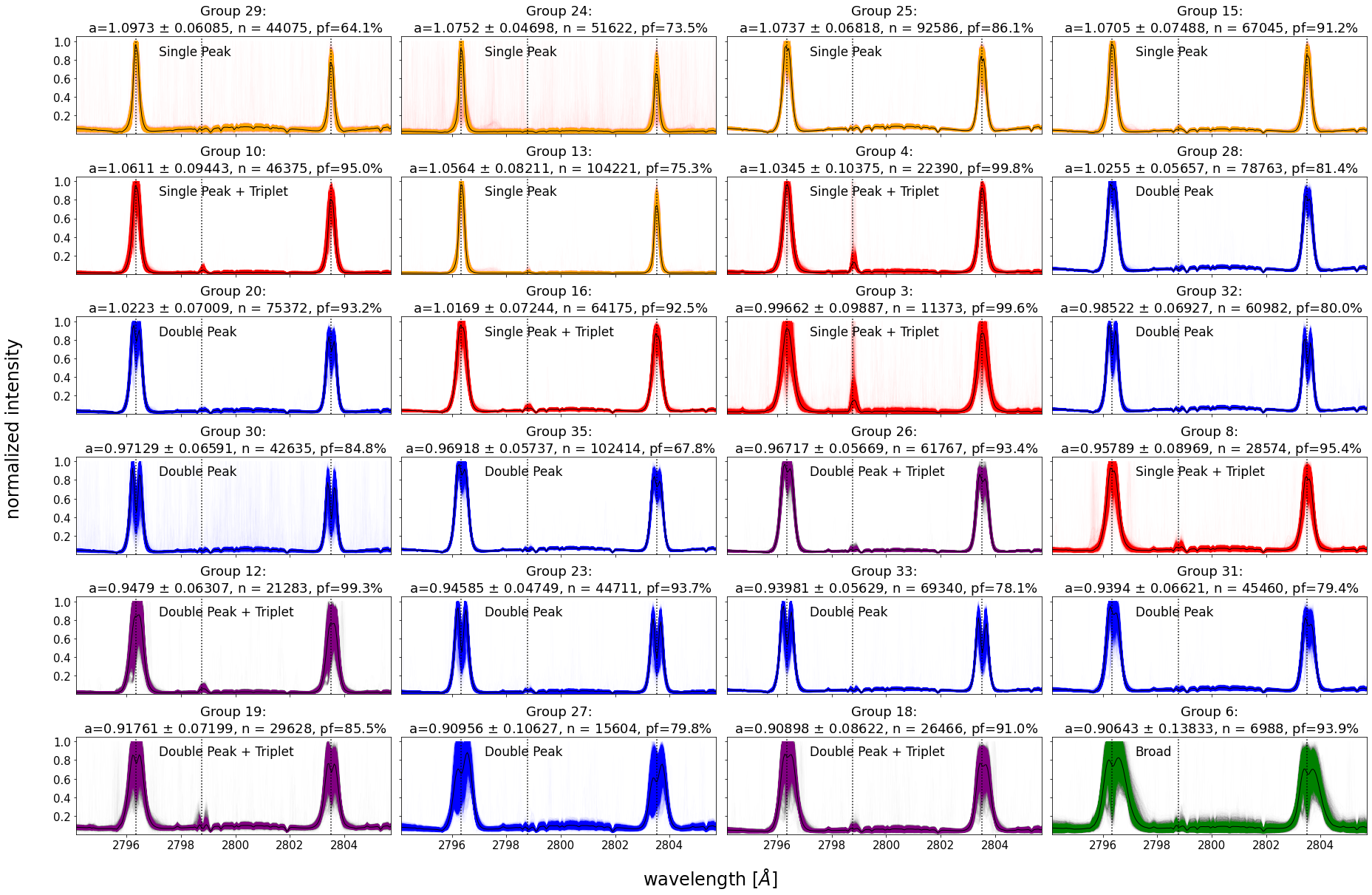}  
  \end{center}
  \caption{Top 24 spectral groups in terms of mean normalized abMIL attention ($\gamma=1$), sorted by mean attention in descending order. For each group the mean attention (a) and standard deviation ($\pm$), the number of assigned spectra (n) and what percentage of the assigned spectra originates from pre-flare observations (pf) are indicated. In addition to the group centroid (black), a sample of at maximum 10'000 assigned spectra is visualized (in color). The groups of spectra are the same as in the ibMIL case in Fig.~\ref{fig:characteristic_groups_ibmil}, however the sorting is different. It is apparent that abMIL's focus is on spectra from both pre-flare and non-flaring active regions.}
  \label{fig:characteristic_groups_abmil}
\end{figure*}

\end{appendix}

\end{document}